\documentclass{emulateapj}
\usepackage{ifpdf}
\usepackage{natbib}
\usepackage{graphicx}
\usepackage{multirow}
\usepackage{color}
\usepackage{cancel}
\usepackage{hyperref}
\usepackage{amsmath}
\usepackage[normalem]{ulem}

\advance \voffset by 1.00cm\relax

\def\msun{{\rm ~M}_{\odot}}
\def\rsun{{\rm ~R}_{\odot}}

\def\zsun{{\rm Z}_{\odot}}
\def\mc{{{\cal M}_c}}
\newcommand\unit[1]{\, {\rm #1}}

\def\be{\begin{equation}}
\def\ee{\end{equation}}
\def\beq{\begin{eqnarray}}
\def\eeq{\end{eqnarray}}
\def\ben{\begin{enumerate}}
\def\een{\end{enumerate}}
\def\bi{\begin{itemize}}
\def\ei{\end{itemize}}
\def\f{\frac}

\usepackage{color}

\newcommand{\ForInternalReference}[1]{{}}

\begin{document}

\title{Double compact objects III: Gravitational-wave detection rates}

 \author{Michal Dominik\altaffilmark{1}, 
         Emanuele Berti\altaffilmark{2}, 
         Richard O'Shaughnessy\altaffilmark{3}, 
         Ilya Mandel\altaffilmark{4}, 
         Krzysztof Belczynski\altaffilmark{1,5},
         Christopher Fryer\altaffilmark{6}, 
         Daniel E. Holz\altaffilmark{7}, 
         Tomasz Bulik\altaffilmark{1}, 
         Francesco Pannarale\altaffilmark{8}
 }

 \affil{
     $^{1}$ Astronomical Observatory, University of Warsaw, Al.
            Ujazdowskie 4, 00-478 Warsaw, Poland  \\
     $^{2}$ Department of Physics and Astronomy, The University of Mississippi, 
	    University, MS 38677, USA \\
     $^{3}$ Center for Gravitation, Cosmology, and Astrophysics, University of Wisconsin-Milwaukee, Milwaukee, WI \\
          $^{4}$ School of Physics and Astronomy, University of Birmingham, Edgbaston, Birmingham B15 2TT, United Kingdom \\
     $^{5}$ Center for Gravitational Wave Astronomy, University of Texas at
            Brownsville, Brownsville, TX 78520\\
     $^{6}$ CCS-2, MSD409, Los Alamos National Laboratory, Los Alamos, NM 87545 \\
     $^{7}$ Enrico Fermi Institute, Department of Physics, and Kavli Institute
for Cosmological Physics\\University of Chicago, Chicago, IL 60637\\
     $^{8}$ School of Physics and Astronomy, Cardiff University, The Parade, Cardiff CF24 3AA, United Kingdom\\
 }

\begin{abstract}
  The unprecedented range of second-generation gravitational-wave (GW)
  observatories calls for
  refining the predictions of potential sources and detection rates. The
  coalescence of double compact objects (DCOs)---i.e., neutron
  star-neutron star (NS-NS), black hole-neutron star (BH-NS), and black
  hole-black hole (BH-BH) binary systems---is the
  most promising source of GWs for these detectors.
  We compute detection rates of coalescing DCOs in
  second-generation GW detectors using the latest models for their
  cosmological evolution, and implementing inspiral-merger-ringdown (IMR)
  gravitational waveform models in our signal-to-noise ratio
  calculations. We find that: (1) the inclusion of the merger/ringdown
  portion of the signal does not significantly affect rates for NS-NS
  and BH-NS systems, but it boosts rates by a factor $\sim 1.5$ for
  BH-BH systems; (2) in almost all of our models BH-BH systems yield
  by far the largest rates, followed by NS-NS and BH-NS systems, respectively, 
  and (3) a majority of the detectable BH-BH systems were formed in the early 
  Universe in low-metallicity environments. We make predictions for the 
  distributions of detected binaries and discuss what the first GW detections will teach us about the
  astrophysics underlying binary formation and evolution.
\end{abstract}


\section{Introduction}

Nearly a century has passed since Albert Einstein wrote down the field
equations of general relativity. A crucial prediction of his theory is
the existence of GWs.  Observations of the Hulse-Taylor binary pulsar
\citep{Taylor:1989} and the double pulsar J0737-3039 \citep{Lyne:2004}
leave little doubt of the existence of GWs, with further evidence
provided by the recent claim of a detection of a GW-induced B-mode
polarization of the cosmic microwave
background~\citep{2014arXiv1403.3985B}.  However, GWs still elude
direct observation.  The situation should change in the next few
years, when a network of second-generation GW
observatories -- including Advanced LIGO (Harry, \citeyear{AdvLIGO},
henceforth aLIGO),
Advanced Virgo \citep[][henceforth AdV]{AdvVirgo}, and KAGRA
\citep{KAGRA} -- will start taking data.  The unprecedented
sensitivity of these observatories will allow them to observe the
inspiral and merger of DCOs out to cosmological distances: for
example, aLIGO should observe binary neutron stars out to a luminosity
distance of $\simeq 450 \unit{Mpc}$ ($z \sim 0.1$), while DCOs
containing BHs will be observable to much larger distances
\citep[e.g.,][]{2010CQGra..27q3001A}.  Given the cosmological reach of
second-generation GW interferometers, a theoretical investigation of
the observable DCO populations which incorporates cosmological
evolution and accurate models of the gravitational waveforms is
particularly timely. This is the goal of this paper, the third in a
series \cite[cf.][]{dominik,dominik2}. Our work builds on the results
presented in the second paper \citep[henceforth Paper 2]{dominik2},
where we presented the cosmological distribution of DCOs for a set of
four evolutionary models.  These models investigated a range of
Hertzsprung gap (HG) common envelope (CE) donors, supernova (SN)
explosion engines, and BH natal kicks, showing distinct differences in
the properties of the resulting DCO populations. Population models
were placed in a cosmological context by adopting the star formation
history reported in \cite{strolger} and the galaxy mass distribution
of \cite{fontana}, both of which are redshift-dependent. We performed
all calculations assuming two scenarios for metallicity evolution,
meant to bracket the uncertainties associated with the chemical
composition of the Universe. Binary evolution was performed using the
{\tt StarTrack} population synthesis code \citep{startrack}.

In this work we complete and extend the analysis of Paper 2. We study
the detection rates and the expected physical properties of coalescing
DCOs at cosmological distances for second-generation GW observatories.
The rates are calculated for different sets of gravitational waveform
models and different detector sensitivities, representative of aLIGO,
AdV, and KAGRA.  Several different groups have presented similar
estimates and studies in the past decade
\citep[e.g.,][]{lipunov1997,bethe,dedonder,bloom,grishchuk:2001,nele2001,voss,dewi,nutzman,pfahl,
danny,PostnovYungelson:2006,seba,mennekens}. However,
none have combined cosmological DCO populations with accurate GW
models to obtain thorough, detector-specific results.
Our astrophysical models for DCO formation are reviewed in Section
\ref{binevol}. Gravitational waveform models and signal-to-noise ratio
estimates are discussed in Section \ref{wmodels}. Our procedure to
compute event rates is presented in Section \ref{sec:fullrates}. Event
rates and bulk properties of the detected populations are presented in
Section \ref{sec:results}. In Section \ref{sec:nobhbh} we present and discuss the study 
by \cite{mennekens}, the primary result of which is the lack of detectable BH-BH systems. 
In Section \ref{sec:conclusions} we discuss the possible astrophysical payoff of the 
first GW detections and important directions for future work.

\section{Astrophysical models} \label{binevol}

\subsection{Binary evolution} 

We begin with a summary of the four {\tt StarTrack}
evolutionary models that form the backbone of this work; a more
detailed discussion can be found in \cite{dominik,dominik2}.

\smallskip
\noindent
\textit{1) Standard model}. This is our reference model, representing
the state of the art in the formation and evolution of binary
systems. We consider only field populations here. Rate estimates
performed for dense populations in which dynamical interactions
between stars are important (i.e., globular clusters and galactic nuclear clusters) have been
presented elsewhere
\citep{gultekin,oleary,grin2006,sadowski,ivan,downing,MillerLauburg:2008}.  Our
Standard model uses the ``Nanjing'' \citep{chlambda} $\lambda$
coefficient in the CE energy balance prescription of \cite{webbink},
where the precise value of $\lambda$ depends on the evolutionary stage
of the donor, its Zero Age Main Sequence (ZAMS) mass, the mass of its
envelope, and its radius. In turn, these quantities depend on
metallicity, which in our simulations varies within the broad range
$10^{-4}\leq Z \leq 0.03$ (recall that solar metallicity corresponds
to $\zsun=0.02$). The values of $\lambda$ for high-mass stars
($M_{ZAMS}>20\msun$) were obtained through private communication with
the authors and are not present in \cite{chlambda}.

The impact of the CE outcome on binary populations depends strongly on the 
evolutionary stage of the donor, as first discussed in \cite{rarity}. The 
Standard model does not allow for CE events with HG donors. These stars are 
not expected to possess a clear core-envelope structure \citep{ivanovataam}, 
thus making it difficult for them to eject their outer layers during the CE phase. 
In our Standard model all CE events with HG donors lead to a prompt
merger before a DCO binary is formed, regardless of the 
aforementioned energy balance.

The model employs a Maxwellian distribution of natal kicks for NSs
with 1-D root mean square velocity $\sigma=265$ km/s, consistent with
NS observations \citep{hobbs}. The same distribution is extended to
BHs, where we allow for the possibility that the kicks may be reduced due
to fallback of material during the SN that leads to BH formation. The
reduction in BH kicks is described via
\begin{equation} \label{vkick}
V_{\rm k}=V_{\rm max}(1-f_{\rm fb}),
\end{equation}
where $V_{\rm k}$ is the final magnitude of the natal kick, $V_{\rm max}$ 
is the velocity drawn from a Maxwellian kick distribution,
and $f_{\rm fb}$ is a ``fallback factor'' that depends on the amount 
of fallback material, calculated according to the prescription given in
\cite{chrisija}. 
Our Standard model uses the ``Rapid'' convection-driven, neutrino-enhanced SN engine \citep{chrisija}. 
The SN explosion is sourced from the Rayleigh-Taylor instability and occurs within the first
$0.1\,$--$\,0.2\,\mbox{s}$ after the bounce. When used in the context of binary evolution
models, this SN engine successfully reproduces the mass gap
\citep{massgap} observed in Galactic X-ray binaries \citep{mg1,mg2},
but see also \cite{2012ApJ...757...36K}.

\smallskip
\noindent
\textit{2) Optimistic Common Envelope}. In this model we allow HG
stars to be CE donors. When the donor initiates the CE phase, the CE
outcome is determined via energy balance. The remaining physics is
identical to the Standard model.

\smallskip
\noindent
\textit{3) Delayed SN}. This model utilizes the ``Delayed'' SN engine
instead of the Rapid one. The former is also a convection driven,
neutrino enhanced engine, but is sourced from the standing accretion
shock instability (SASI), and can produce an explosion as late as
$1\,\mbox{s}$ after bounce. The Delayed engine produces a continuous
mass spectrum of compact objects, ranging from NSs through light BHs to
massive BHs \citep{massgap}.

\smallskip
\noindent
\textit{4) High BH kicks}. In this model the BHs receive full natal
kicks, i.e. we set $f_{\rm fb}=0$ in Eq.~(\ref{vkick}). Otherwise this model is
identical to the Standard model.
 
\subsection{Metallicity evolution}\label{sec:metallicity}

In this paper we employ two distinct metallicity evolution scenarios:
``high-end'' and ``low-end''. These are
identical to those in our previous study (Paper 2), and a detailed 
description can be found therein. 
Employing such calibrations allows us to explore and bracket
uncertainties in the chemical evolution of the Universe.
In both cases the average metallicity decreases with increasing
redshift.

The high-end metallicity profile is calibrated to yield a median value
of metallicity equal to $1.5\,\zsun$ (or $8.9$ in the ``12+log(O/H)''
formalism) at redshift $z=0$. This calibration was designed to match
the upper $1 \sigma$ scatter of metallicities according to \cite{yuan}
(see their Fig.~2, top-right panel).

The low-end metallicity profile is based on SDSS observations \citep{panter},
from which we infer that one half of the star forming mass of galaxies
at $z\sim0$ has $20\%$ solar metallicity, while the other half has
$150\%$ solar metallicity. 
\section{Waveform models} \label{wmodels}

\subsection{Order-of-magnitude estimates} \label{sec:simplerates}

\begin{figure}
\includegraphics[width=1.0\columnwidth,clip=true]{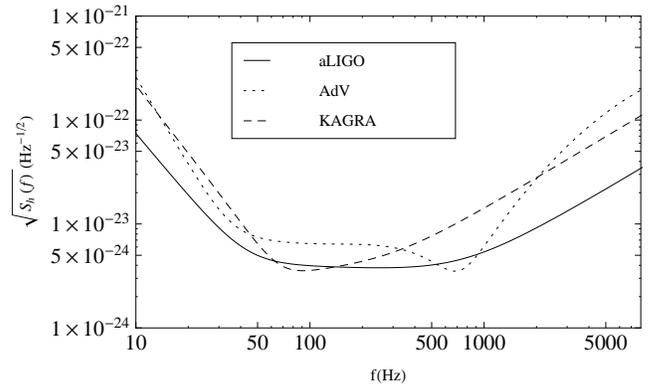}
\caption{\label{fig:noise} \textbf{Noise models}: we use an analytical approximation to the aLIGO zero-detuning
  high power (ZDHP) noise power spectral density given in Eq.~(4.7) of \cite{ajithspin}
  (we verified that this approximation gives results in excellent
  agreement with the ``official'' tabulated aLIGO ZDHP noise
  PSD given in \cite{PSD:AL}. For AdV we use the fit in Eq.~(3.4) of
  \cite{ajithbose} to \cite{AdvVirgo}, and for KAGRA we use the PSD fit from the Appendix of
  \cite{pannarale} to \cite{KAGRA}.}
\end{figure}

For any given GW detector the ``horizon distance'',  $D_h$,
is defined as the luminosity distance at which an optimally oriented (face-on, overhead) 
canonical $(1.4+1.4)~M_\odot$ NS-NS binary would be detected at a 
fiducial threshold signal-to-noise ratio (SNR), taken to be $8$ in this paper. The 
expectation value of the SNR, $\rho$, of a signal with GW amplitude $h(t)$ is given by
\be\label{SNR}
\rho^2 = 4\int_0^\infty \f{|\tilde h(f)|^2}{S_n(f)} df\,,
\ee
where $\tilde h(f)$ is the Fourier transform of the signal and
$S_n(f)$ is the noise power spectral density of the detector
\citep[see e.g.][]{cutlerflanagan,poissonwill}. The square root of the
noise power spectral density is plotted in Fig.~\ref{fig:noise} for
several advanced interferometers of interest. For example, the aLIGO
horizon distance is $D_{h} \simeq 450 \unit{Mpc}$.

Although the sensitivity of a GW detector network depends 
on the details of the search pipeline and the detector data quality, we 
follow \cite{2010CQGra..27q3001A} in considering a single detector with an 
SNR threshold $\rho \ge 8$ as a proxy for detectability by the network.  
With this criterion, a simple and common expression to transform the local 
merger rate to a predicted detection rate $R_D$, given the horizon distance
$D_h$ and the merger rate density, ${\cal R}(z)$, evaluated locally
(at $z=0$), is:
\begin{equation}
\label{eq:LocalUniverseMergerRateFormula}
R_D \simeq  \frac{4\pi}{3} D_{h}^3 \langle w^3\rangle \left<(\mc/1.2 M_\odot)^{15/6}\right>{\cal R}(0) 
\end{equation}
In this expression $\langle w^3\rangle^{-1/3}\simeq 2.264$ is a
purely geometrical and SNR-threshold-independent factor commonly used to relate sky 
location- and orientation-averaged distances to optimal detection distances (see Appendix 
for details) and $\mc=\eta^{3/5}M$ (where $M=m_1+m_2$ is the total mass of
the binary and $\eta\equiv m_1m_2/M^2$) is the ``chirp mass'' \citep[see, e.g.,][]{cutlerflanagan}.
This estimate assumes that (1) cosmological effects are negligible
(i.e., space is Euclidean to a good approximation), and (2) most of
the SNR is accumulated during an inspiral phase which lasts through
the entire sensitive band of the detector, where the GW amplitude in
the frequency domain is well approximated by the quadrupole formula,
i.e., $\tilde h(f)\sim \mc^{5/6} f^{-7/6}/D$. Here $D$ is the
luminosity distance to the source. The estimate of
Eq.~(\ref{eq:LocalUniverseMergerRateFormula}) follows from this simple
scaling together with the definition of the SNR, Eq.~(\ref{SNR}).

Eq.~(\ref{eq:LocalUniverseMergerRateFormula}) involves only the
\emph{local} merger rate ${\cal R}(0)$ and $\langle \mc^{15/6}\rangle$
is averaged over detected binaries. Both quantities can easily be
extracted from {\tt StarTrack} simulations; they are listed in Table
\ref{tab:simplerates}, along with the values of $R_D$ predicted by
Eq.~(\ref{eq:LocalUniverseMergerRateFormula}). We expect this rough
estimate to be accurate for NS-NS binaries, for which the overwhelming
majority of the SNR is accumulated during the inspiral phase.  More
accurate calculations are required for DCOs comprised of BHs, because
they are visible out to larger distances (making cosmological
corrections important) and because, as we discuss below, a large
fraction of the SNR for these binaries comes from the merger/ringdown
portion of the signal.

\subsection{Including merger and ringdown} \label{sec:IMR}

In order to refine our rate estimates for high-mass systems containing
BHs, it is important to consider the full waveform, including
inspiral, merger, and ringdown (IMR). The calculation of gravitational
waveforms from merging BH-BH and BH-NS binaries requires expensive
numerical relativity simulations, but several semi-analytical models
have been tuned to reproduce the amplitude and phasing of BH-BH and
BH-NS merger simulations. To estimate systematic uncertainties and the
impact of spin, we performed rate calculations using three models: (1)
the IMRPhenomB model described in \cite{PhenomB}, one of the earliest
phenomenological models tuned to both nonspinning and spinning BH-BH
simulations with aligned spins, henceforth abbreviated as PhB; (2) the
IMRPhenomC (henceforth abbreviated PhC) model by \cite{santamaria}, a
more accurate alternative to PhB also tuned to nonprecessing
simulations of BH-BH mergers; and (3) a nonspinning effective-one-body
(EOB) model \citep{eob}. A detailed comparison of the three models can
be found in \cite{Damour:2010zb}.
Recent work by \cite{pannarale} shows that finite-size effects
introduce negligible errors ($\lesssim 1\%$) in SNR calculations for
BH-NS binaries, therefore the above models are adequate for {\em both} BH-BH
and BH-NS binaries.
In order to facilitate comparison with previous work, we also
evaluated rates using the simplest possible approximation: a
restricted post-Newtonian (PN) waveform where the amplitude is
truncated at Newtonian order, i.e. $\tilde h(f)\sim \mc^{5/6}
f^{-7/6}/D$, terminated at a fiducial ``innermost stable circular
orbit'' frequency $f_{\rm ISCO}=(G M\pi/c^3)^{-1}6^{-3/2}$. At low
mass, the upper limit can be neglected and this approximation
corresponds to $\rho \propto \mc^{5/6}$, as stated above: see also
Eq.~(7) in \cite{roskb}.

\begin{figure}
\includegraphics[width=\columnwidth]{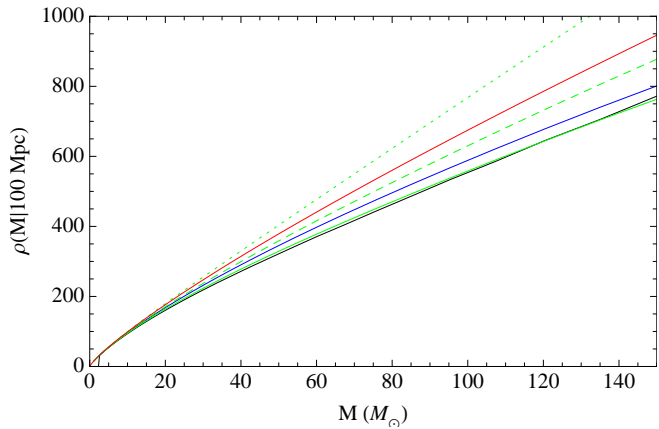}
\caption{\label{fig:Ingredients:SNRVersusMass:CompareModels}\textbf{SNR
    for different signal models}: To illustrate the relatively small
  differences between the signal models we have adopted, we show the
  SNR, $\rho(M)$, as a function of total binary mass, $M$, for an
  equal-mass nonspinning binary at $100 \unit{Mpc}$, where the SNR is
  evaluated using a single fiducial aLIGO detector.  The colored solid
  curves show (a) the trivial expression $\rho =\rho_0(M/2.8
  M_\odot)^{5/6}$ with $\rho_0=34.3$ (red), (b) an EOB model (black),
  PhB model (blue), and PhC model (green), all evaluated for zero
  spin. The green dotted line shows the PhC model evaluated with
  near-extremal spin on both objects ($\chi_1=\chi_2=0.998$), while
  the green dashed line shows PhC with near-extremal spin on one
  object ($\chi_1=0.998,\chi_2=0$). The choice $\chi_i=0.998$
  corresponds to the \cite{Thorne:1974ve} bound. This value of the
  spin is outside the regime in which phenomenological models have
  been calibrated, and it has been chosen to provide rough upper
  limits on the rates.}
\end{figure}

Figure \ref{fig:Ingredients:SNRVersusMass:CompareModels} shows that
these models all make similar predictions for the SNR of optimally
oriented equal-mass binaries as a function of their total mass for a
single aLIGO detector.  Even small differences can be important: for
any given binary, a $30\%$ difference in amplitude corresponds to a
factor $(1.3)^3\simeq 2.2$ in rate calculations. In practice, however,
all nonspinning IMR models agree in SNR to within tens of percent over
the total binary mass range of interest (up to $127\msun$, see
Section~\ref{dcos}). The effect of spin will be discussed in more
detail in Section~\ref{subsec:WF} below.

\begin{figure*}
\ifpdf{
\includegraphics[width=1.0\columnwidth,clip=true]{AdLIGOZDHP_20Hz_SNR100Mpc_ins}
\includegraphics[width=1.0\columnwidth,clip=true]{AdLIGOZDHP_20Hz_SNR100Mpc_IMR}
}
\else
\includegraphics[width=1.0\columnwidth,clip=true]{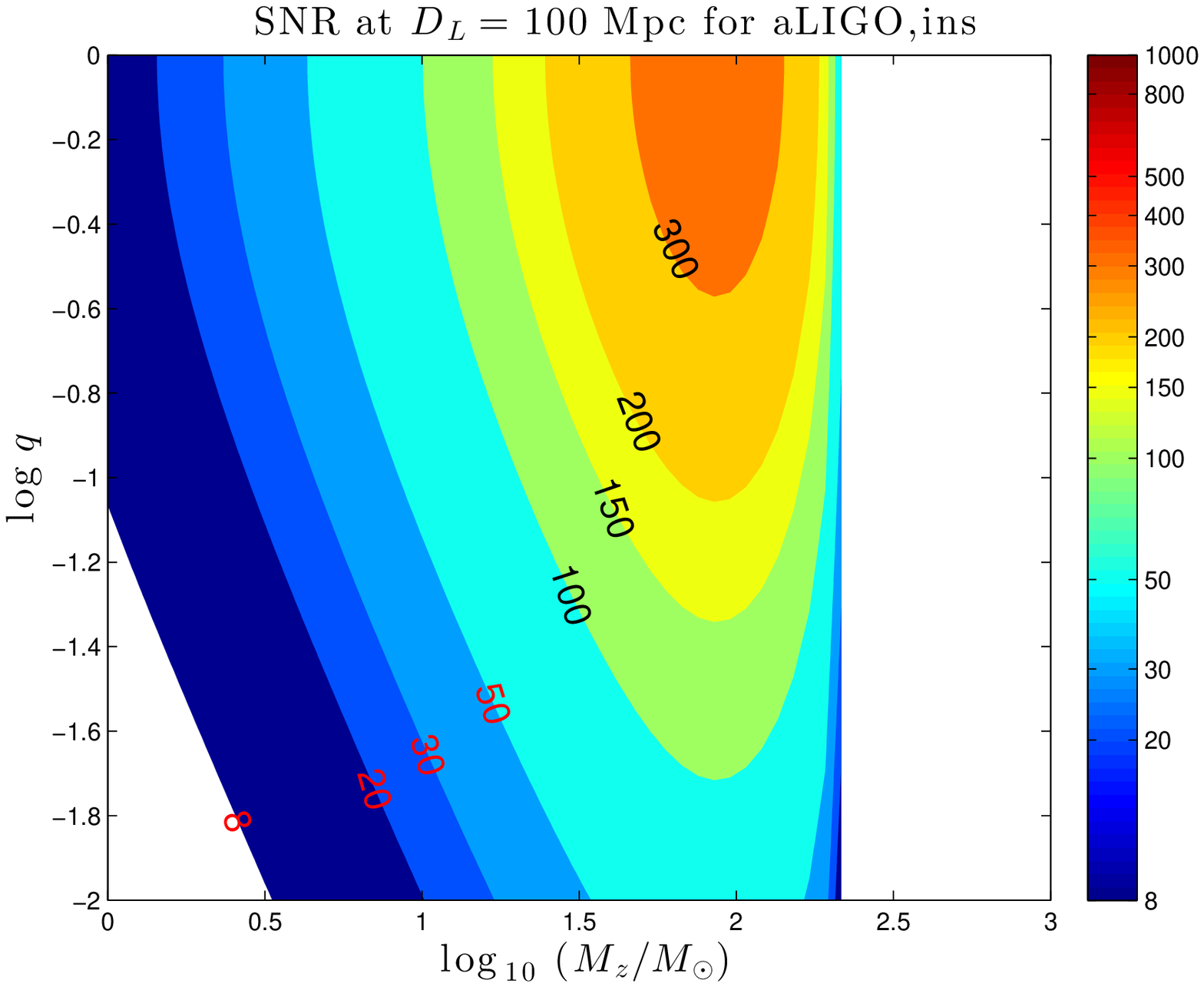}
\includegraphics[width=1.0\columnwidth,clip=true]{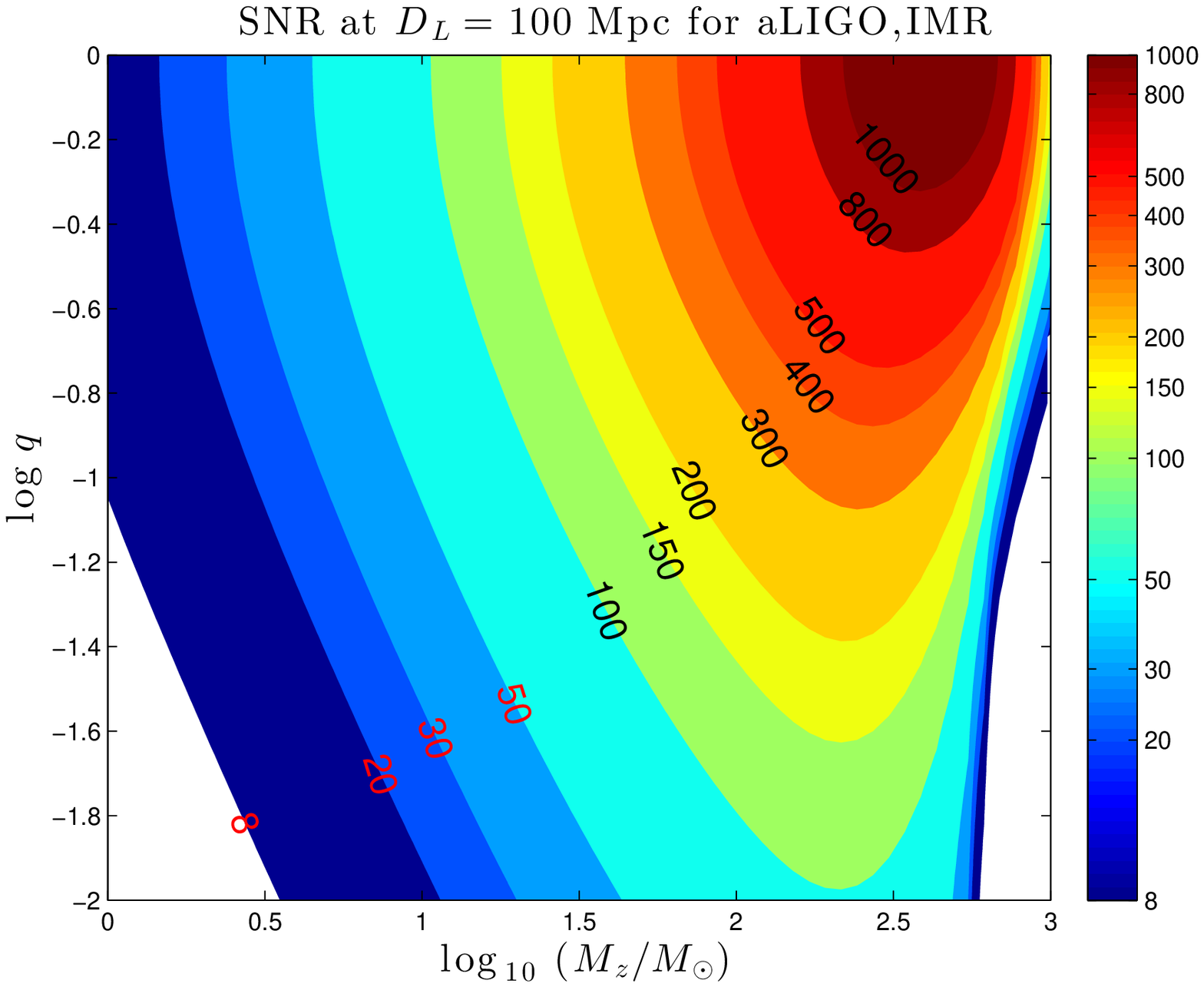}
\fi
\caption{\label{fig:Ingredients:SNRContours}\textbf{Optimal SNR for
  nonspinning binaries} of given (redshifted) total mass $M_z=M(1+z)$
  and mass ratio $q=m_2/m_1$ at luminosity distance $D_L=100$~Mpc. In
  the left panel the SNR is computed using the restricted PN
  approximation (i.e., the GW amplitude is evaluated using the
  quadrupole formula). In the right panel we use the PhC model for the
  full IMR waveform; the results for the EOB model are very similar.
  A low-frequency cutoff of $f_{\rm cut}=20$Hz has been assumed (see
  Section \ref{subsec:WF} for further details).}
\end{figure*}

\begin{figure*}
\ifpdf{
\includegraphics[width=1.0\columnwidth,clip=true]{AdLIGOZDHP_20Hz_DLhor_ins}
\includegraphics[width=1.0\columnwidth,clip=true]{AdLIGOZDHP_20Hz_DLhor_IMR}\\
}
\else
\includegraphics[width=1.0\columnwidth,clip=true]{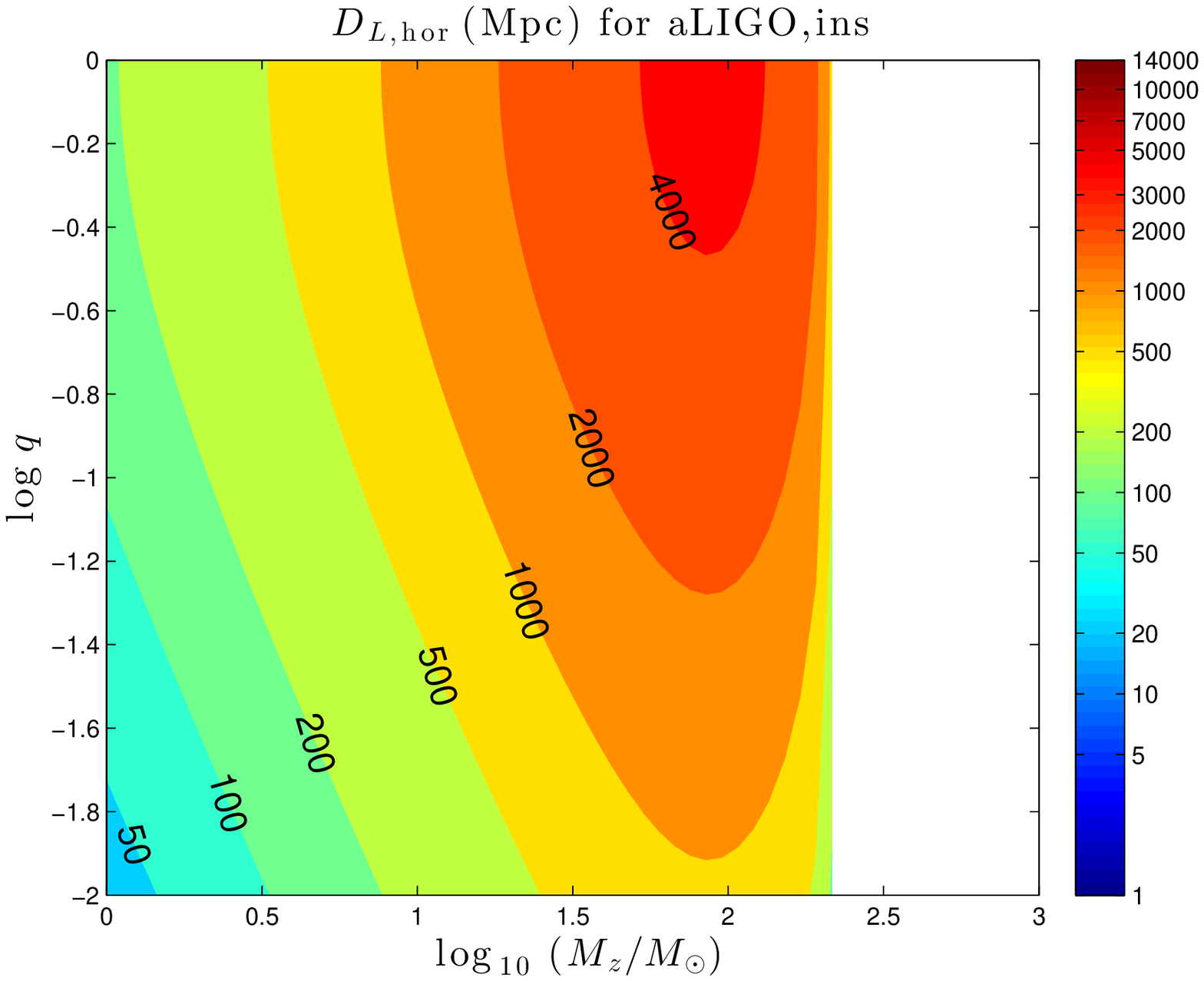}
\includegraphics[width=1.0\columnwidth,clip=true]{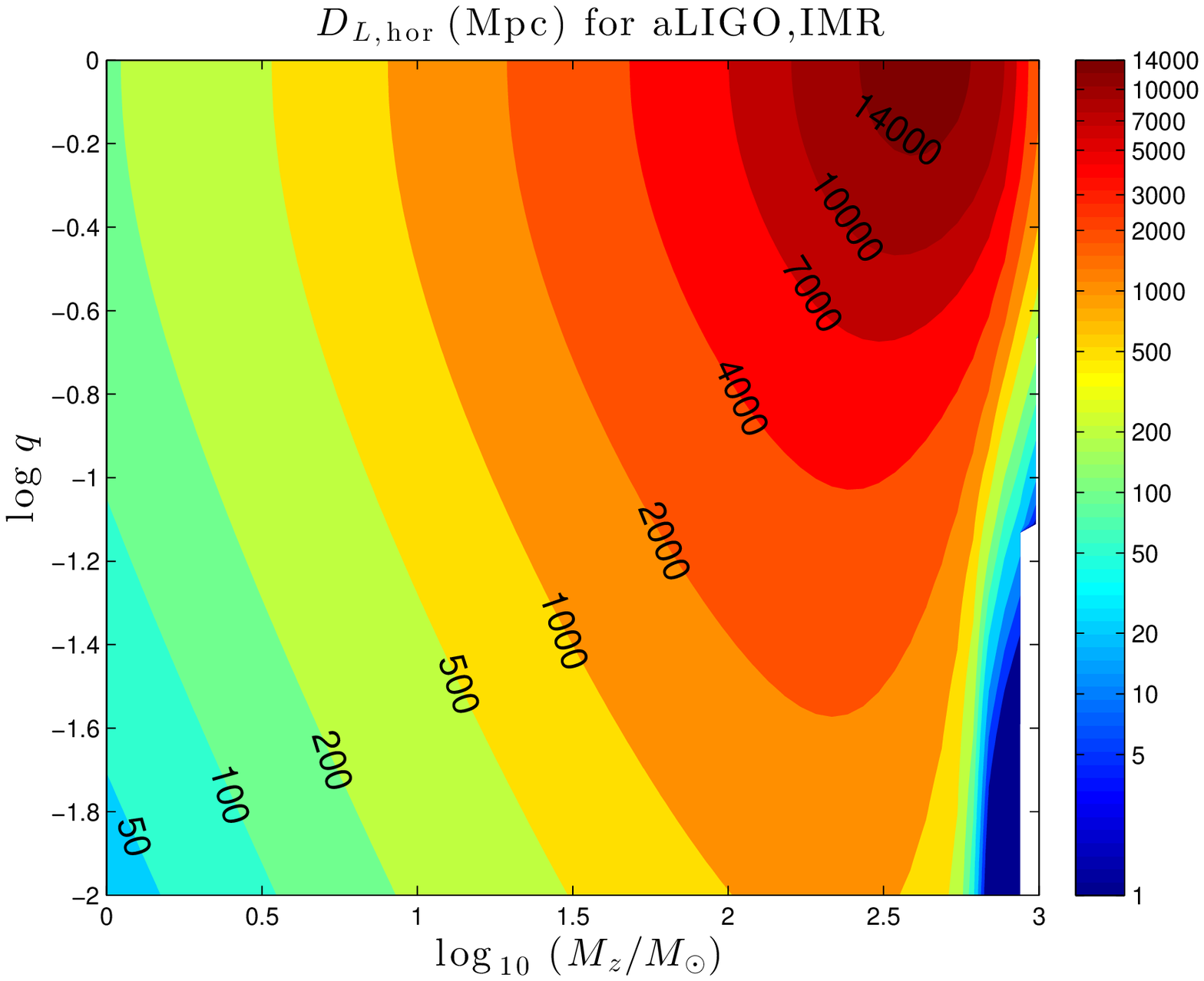}\\
\fi
\caption{\label{fig:Ingredients:Dhorizon}\textbf{Horizon luminosity distance
  (in Mpc) for nonspinning binaries} as a function of redshifted mass
  and mass ratio, computed according to Eq.~(\ref{DLhor}) using
  waveforms comprised of only the inspiral (left panel) or the full
  IMR signal (right panel).  }
\end{figure*}

Figure \ref{fig:Ingredients:SNRContours} shows contour plots of the
SNR, $\rho$, in the $(M_z, q)$ plane, where $M_z \equiv M(1+z)$ is the
redshifted total mass, $z$ is the redshift, and $q=m_2/m_1 \leq 1$ is
the mass ratio of the components, for nonspinning binaries at
luminosity distance $D_L=100$~Mpc.  We discuss the justification for
considering the SNR as a function of $M_z$ below, but since the chosen
distance corresponds to a negligible redshift $z\simeq 0.023$ using
the cosmological parameters $\Omega_M=0.3$, $\Omega_\Lambda=0.7$,
$\Omega_{\rm k}=0$, and $h=0.7$ (chosen for consistency with
\cite{dominik,dominik2}), $M\simeq M_z$ at this distance.  The left
panel refers to a calculation using an inspiral-only waveform with
Newtonian amplitude to compute the horizon distance.  The right panel
includes inspiral, merger, and ringdown, modelled using the PhC
waveform.  This plot shows two important features: (1) including the
full IMR increases the maximum SNR at this luminosity distance by
factors of a few with respect to an inspiral-only calculation, from
$\approx 300$ to $\approx 10^3$; (2) high-mass binaries ($M_z \gtrsim
10^{2.5}M_\odot \approx 300 M_\odot$) involving BHs that would not be
detectable using inspiral waveforms, become detectable using IMR
waveforms. The latter point is not important for the field binaries
considered in this paper, but it is crucial for intermediate-mass BH
mergers \citep[e.g.,][]{walczak,Fregeau:2006,Amaro:2006imbh}.

In an expanding Universe, GW emission is redshifted by the same factor
of $(1+z)$ as electromagnetic radiation. In the units ($G=c=1$)
adopted by relativists to describe gravitational waves, the only
quantity with dimensions in the GW signal is the total mass $M$. Since
the total mass sets the time scale, a binary source of mass $M$ in the
local universe has an identical waveform (but with different
amplitude) to a binary at redshift $z$ with mass $M/(1+z)$; see, e.g.,
\cite{1998PhRvD..57.4535F}.  Eq.~(\ref{SNR}), together with the fact
that gravitational amplitudes scale inversely with the luminosity
distance $D_L(z)$, implies that the horizon redshift $z_{\rm h}$
(i.e., the redshift at which an optimally located and oriented binary
would have SNR $\rho_{\rm thr}=8$) can be found via the simple scaling
\be\label{DLhor}
D_h (z_{\rm h})  = D_L(z) \frac{\rho(M_z, q)} {\rho_{\rm thr}}\,,
\ee
where $\rho$ is the SNR at any redshift $z$, or luminosity distance
$D_L(z)$.  Note that the right-hand side depends only on $z$, $M_z$
and $q$. Therefore one can easily turn an SNR calculation at fixed $z$
(cf. Fig.~\ref{fig:Ingredients:SNRContours}) into a plot of the
horizon luminosity distance $D_h$ (or equivalently of the horizon
redshift $z_h$) such as Fig.~\ref{fig:Ingredients:Dhorizon}.

{\tt StarTrack} produces large catalogs of DCOs with intrinsic
parameters $(M,q)$, with each of these binaries merging at a different
redshift. Any of these representative DCOs is potentially detectable
(depending on precise sky location and binary orientation) when
$z<z_{\rm h}$. Determining detectability therefore amounts to a simple
interpolation of two-dimensional grids similar to those plotted in
Fig.~\ref{fig:Ingredients:Dhorizon}. These grids can be computed once
and for all, given a waveform model and a detector's power spectral
density (PSD).  Evaluating such a grid typically involves $100\times
100=10^4$ SNR evaluations, and it is much faster than the
(impractical) evaluation of millions of SNR integrals such as
Eq.~(\ref{SNR}), one for each representative binary produced by {\tt
  StarTrack}. The conversion between detectability at optimal location
and optimal orientation and detectability at generic orientations
involves a simple geometrical factor $p_{\rm det}$, as discussed
below.

\section{Rate calculation} \label{sec:fullrates}

The detection rate is 
\beq \label{Rdet}
R_{\rm det} = \iiint_0^\infty {\cal R} (z_m) \frac{dt_m}{dt_{\rm det}} p_{\rm det} \frac{dV_c}{dz_m} dz_m dm_1 dm_2,
\eeq
where ${\cal R} (z_m) \equiv \frac{dN}{dm_1 dm_2 dV_c dt_m}$ is the binary merger rate per 
unit component mass per unit comoving volume $V_c$ per unit time $t_m$ as measured in the 
source frame at merger redshift $z_m$, the term 
$\frac{dt_m}{dt_{\rm det}} = \frac{1}{1+z_m}$ accounts for the difference in clock rates 
at the merger and at the detector, and $p_{\rm det}=p_{\rm det} (z_m; m_1, m_2)$ is the probability (over isotropic sky locations and orientations) that a source with given masses at a given redshift will be detectable. The quantity
\be 
\f{dV_c}{dz}=\f{4\pi c}{H_0}\f{D_c^2(z)}{E(z)}\,,
\ee
with $E(z)=\sqrt{\Omega_{\rm M}(1+z)^3+\Omega_{\rm k}(1+z)^2+\Omega_{\Lambda}}$, 
is the comoving volume per unit redshift, and 
\be
D_c(z)=\f{c}{H_0}\int_0^z \f{dz'}{E(z')}
\ee
is the comoving distance, related to the luminosity distance $D_L(z)$
by $D_c(z)=D_L(z)/(1+z)$: see \cite{hogg} for our notation and
conventions. 

The merger rate ${\cal R} (z_m)$ is a convolution of the star formation rate and the number 
density of binaries per unit star forming mass $M_f$ per unit time delay between formation and merger $\tau$:
\beq \label{Rzm}
{\cal R} (z_m) &=& \int_0^{t_m} \int_0^{t_{\rm det}}  \frac{dM_f}{dV_c dt_f} (z_f) 
\frac{dN} {dM_f dm_1 dm_2 d\tau} (t_f; m_1, m_2, \tau) \nonumber\\
&\times& \delta(t_m-t_f-\tau) d\tau dt_f,
\eeq
where ${\rm SFR} = \frac{dM_f}{dV_c dt_f} (z_f)$ is the star formation rate
per unit comoving volume per unit time $t_f$ at formation redshift $z_f$. 

The distribution of binaries in mass and time delay space, $\frac{dN} {dM_f dm_1
  dm_2 d\tau} (t_f; m_1, m_2, \tau)$, is obtained with the {\tt StarTrack}
population synthesis code, taking into account the metallicity distribution at
the formation redshift as described in Section~\ref{binevol}.  Since {\tt StarTrack} simulations
produce a set of merging binaries with specific component masses and time delays
sampling the desired distribution, the integrals above are easily computed
via Monte Carlo over the simulated systems.  For computational efficiency
the outer integral over the time of formation in Eq.~(\ref{Rzm}) is binned over
$\Delta t_f = 100$ Myr segments, while the integral over the merger redshift
$z_m$ in Eq.~(\ref{Rdet}) is transformed into an integral over merger time via $dz_m
= \frac{dz_m}{dt_m} dt_m = H_0 (1+z)E(z) dt_m$ \citep{hogg}.
Thus the detection rate integral can be represented as a Monte Carlo sum over all simulated binaries:
\be \label{eq:rate}
R_{\rm det}=\sum \frac{\rm SFR}{\Delta M_f} p_{\rm det} \frac{1}{1+z} \frac{dV_c}{dz} \frac{dz}{dt} \Delta t \,, 
\ee 
where $\Delta M_f$ is the total star-forming mass that was simulated
in the Monte Carlo to represent the time bin $\Delta t$, all terms but
the first are computed at the merger redshift of the simulated source.

The detection probability for a given source at its merger redshift
$p_{\rm det} (z, m_1, m_2)$ is simply the fraction of sources of a
given mass located at the given redshift that exceed the detectability
threshold in SNR, assuming that sources are uniformly distributed in
sky location and orbital orientation.  If a single detector with an
SNR threshold (e.g., $\rho_{\rm thr}=8$) is used as a proxy for
detectability, the detection probability can be expressed as a
cumulative distribution function on the projection parameter $w$.  In the Appendix, $w$ is defined such that $w=0$ when the detector has no response to the gravitational wave, and $w=1$ for an optimally located and oriented (face-on and directly overhead) binary. The detection probability is 
\be 
p_{\rm det} = P(\rho_{\rm thr}/\rho_{\rm opt})\,, 
\ee 
where $P(w)$ is the cumulative distribution function on $w$ over different source locations and orientations, and $\rho_{\rm opt}$ is the signal-to-noise ratio (SNR) for an optimally located
and oriented binary at redshift $z$.

\begin{center}
\begin{deluxetable*}{l|ll|ll}[thb]
\tablewidth{330pt} \tablecaption{Local merger rates and simply-scaled detection rate predictions\tablenotemark{a}:}
\startdata
Model   & $\left<\mc^{15/6}\right>$& ${\cal R}(0)$  & $R_D$ (aLIGO $\rho \ge 8$) & $R_D$ (3-det network $\rho
\ge 10$)\\
   & $M_\odot^{15/6}$  & $\unit{Gpc}^{-3} \unit{yr}^{-1}$ & $\unit{yr}^{-1}$ & $\unit{yr}^{-1}$\\\hline
{NS-NS}\\
Standard       & 1.1 (1.1) & 61  (52)  & 1.3 (1.1) & 3.2  (2.7)\\
Optimistic CE  & 1.2 (1.2) & 162 (137) & 3.9 (3.3) & 9.2   (7.7) \\
Delayed SN     & 1.4 (1.4) & 67  (60)  & 1.9 (1.7) & 4.5  (4.0) \\
High BH Kicks  & 1.1 (1.1) & 57  (52)  & 1.2 (1.1) & 3.0  (2.7)  \\
\hline
{BH-NS}\\
Standard       & 18 (19) & 2.8  (3.0)  & 1.0  (1.2)  & 2.4  (2.7)  \\
Optimistic CE  & 17 (16) & 17   (20)   & 5.7  (6.5)  & 13.8   (15.4) \\
Delayed SN     & 24 (20) & 1.0  (2.4)  & 0.5  (0.9)  & 1.1  (2.3)\\
High BH Kicks  & 19 (13) & 0.04 (0.3)  & 0.01 (0.08) & 0.04 (0.2)\\
\hline
{BH-BH} \\
Standard       & 402  (595)  & 28  (36)  & 227 (427)  & 540 (1017)\\
Optimistic CE  & 311  (359)  & 109 (221) & 676 (1585) & 1610 (3773)\\
Delayed SN     & 829  (814)  & 14  (24)  & 232 (394)  & 552 (938)\\
High Kick      & 2159 (3413) & 0.5 (0.5) & 22  (34)   & 51  (81)
\enddata
\label{tab:simplerates}
\tablenotetext{a}{Detection rates computed using the basic scaling of
  Eq.~(\ref{eq:LocalUniverseMergerRateFormula}) for both the
  \textit{high-end} and \textit{low-end} (the latter in parentheses) metallicity
  scenarios (see Section \ref{sec:metallicity}). These rates should be
  compared with those from more careful calculations presented in
  Tables~\ref{rates2genH} and~\ref{rates2genL}.
}
\end{deluxetable*}
\end{center}

\begin{deluxetable*}{l|ll|ll|lll|ll}[thb]
\tablecaption{Detection rates for second-generation detectors in the
  \textit{high-end} metallicity scenario}
\startdata
& \multicolumn{2}{c}{AdV [$\rho \ge 8$]}
& \multicolumn{2}{c}{KAGRA [$\rho \ge 8$]}
& \multicolumn{3}{c}{aLIGO [$\rho \ge 8$]}
& \multicolumn{2}{c}{3-det network [$\rho \ge 10 (12)$]}\\
& \multicolumn{2}{c}{$f_{\rm cut}=20$~Hz}
& \multicolumn{2}{c}{$f_{\rm cut}=10$~Hz}
& \multicolumn{3}{c}{$f_{\rm cut}=20$~Hz}
& \multicolumn{2}{c}{$f_{\rm cut}=20$~Hz}\\
Model   &Insp   &PhC (EOB) &Insp   &PhC (EOB) &Insp   &PhC (EOB) & PhC (spin) & Insp & PhC\\
   & $\unit{yr}^{-1}$ & $\unit{yr}^{-1}$ & $\unit{yr}^{-1}$ & $\unit{yr}^{-1}$
   & $\unit{yr}^{-1}$ & $\unit{yr}^{-1}$ & $\unit{yr}^{-1}$ & $\unit{yr}^{-1}$ & $\unit{yr}^{-1}$ \\\hline
{NS-NS}\\
Standard       & 0.3 & 0.3 & 0.8 & 0.7 & 1.2 & 1.1 &- & 2.5 (1.5) & 2.4 (1.4) \\
Optimistic CE  & 0.9 & 0.9 & 2.1 & 1.9 & 3.3 & 3.1 &- & 6.9 (4.0) & 6.5 (3.8) \\
Delayed SN     & 0.4 & 0.4 & 1.0 & 0.9 & 1.6 & 1.5 &- & 3.3 (1.9) & 3.1 (1.8) \\
High BH Kicks  & 0.3 & 0.3 & 0.7 & 0.7 & 1.1 & 1.1 &- & 2.3 (1.4) & 2.2 (1.3) \\
\hline
{BH-NS}\\
Standard       & 0.2  & 0.2    & 0.5  & 0.4   & 0.7  & 0.6   & 0.8 & 1.5  (0.9) & 1.2 (0.7) \\
Optimistic CE  & 1.1  & 1.0    & 2.9  & 2.2   & 4.4  & 3.6   & 4.4 & 9.2  (5.4) & 7.4 (4.3)    \\
Delayed SN     & 0.09 & 0.07   & 0.2  & 0.2   & 0.4  & 0.3   & 0.5 & 0.8  (0.5) & 0.6 (0.3) \\
High BH Kicks  & 0.01 & 0.007  & 0.02 & 0.02  & 0.04 & 0.03  & 0.1 & 0.09 (0.05) & 0.07 (0.04) \\
\hline
{BH-BH} \\
Standard       & 35  & 41  (38)  & 70  & 93  (86)  & 117 & 148 (142) & 348  & 236  (144) & 306 (177) \\
Optimistic CE  & 126 & 144 (133) & 281 & 366 (333) & 491 & 618 (585) & 1554 & 1042 (588) & 1338 (713) \\
Delayed SN     & 27  & 34  (32)  & 50  & 81  (75)  & 90  & 129 (124) & 320  & 182  (110) & 270 (155) \\
High Kick      & 0.6 & 1.0 (0.9) & 0.9 & 2.5 (2.3) & 2.1 & 3.8 (3.8) & 12   & 4.2  (2.7) & 8.2 (4.7)
\enddata
\label{rates2genH}
\tablenotetext{a}{Detection rates computed for the high-end
  metallicity evolution scenario using the inspiral (``Insp'') and PhC
  or EOB IMR models for nonspinning binaries. For aLIGO we also list
  rough upper limits on the rates computed with the IMR PhC model by
  assuming that BHs have near-maximal aligned spins
  ($\chi_1=\chi_2=0.998$ for BH-BH systems; $\chi_1=0.998$ and
  $\chi_2=0$ for BH-NS systems). The inspiral is calculated using the
  restricted PN approximation, which overestimates the amplitude (and
  therefore the detection rates) for low-mass systems (NS-NS) when
  compared to the full IMR calculations; cf.~Section~\ref{wmodels} for
  details. The last two columns were computed assuming a minimum {\it
    network} SNR of 10 (or 12, in parentheses) for a three-detector
  network composed of three instruments located at the LIGO Hanford,
  LIGO Livingston, and Virgo sites, all with aLIGO sensitivity. For
  each detector, $f_{\rm cut}$ is the assumed low-frequency cutoff in
  the power spectral density: see section \ref{subsec:WF}.}
\end{deluxetable*}

\begin{deluxetable*}{l|ll|ll|lll|ll}[thb]
\tablecaption{Detection rates for second-generation detectors in the
  \textit{low-end} metallicity scenario}
\startdata
& \multicolumn{2}{c}{AdV [$\rho \ge 8$]}
& \multicolumn{2}{c}{KAGRA [$\rho \ge 8$]}
& \multicolumn{3}{c}{aLIGO [$\rho \ge 8$]}
& \multicolumn{2}{c}{3-det network [$\rho \ge 10 (12)$]}\\
& \multicolumn{2}{c}{$f_{\rm cut}=20$~Hz}
& \multicolumn{2}{c}{$f_{\rm cut}=10$~Hz}
& \multicolumn{3}{c}{$f_{\rm cut}=20$~Hz}
& \multicolumn{2}{c}{$f_{\rm cut}=20$~Hz}\\
Model   &Insp   &PhC (EOB) &Insp   &PhC (EOB) &Insp   &PhC (EOB) & PhC (spin) & Insp & PhC \\
   & $\unit{yr}^{-1}$ & $\unit{yr}^{-1}$ & $\unit{yr}^{-1}$ & $\unit{yr}^{-1}$
   & $\unit{yr}^{-1}$ & $\unit{yr}^{-1}$ & $\unit{yr}^{-1}$ & $\unit{yr}^{-1}$ & $\unit{yr}^{-1}$ \\\hline
{NS-NS}\\
Standard       & 0.3 & 0.3 & 0.7 & 0.6 & 1.1 & 1.0 &- & 2.3 (1.3) & 2.2 (1.3) \\
Optimistic CE  & 0.8 & 0.7 & 1.8 & 1.7 & 2.9 & 2.7 &- & 6.0 (3.5) & 5.6 (3.3) \\
Delayed SN     & 0.4 & 0.4 & 1.0 & 0.9 & 1.5 & 1.4 &- & 3.2 (1.8) & 2.9 (1.7) \\
High BH Kicks  & 0.3 & 0.3 & 0.7 & 0.6 & 1.0 & 1.0 &- & 2.1 (1.3) & 2.0 (1.2) \\
\hline
{BH-NS}\\
Standard       & 0.3  & 0.2  & 0.7  & 0.5  & 1.1 & 0.8 & 1.2 & 2.3 (1.3) & 1.8 (1.0) \\
Optimistic CE  & 1.4  & 1.2  & 3.6  & 2.8  & 5.5 & 4.4 & 5.7 & 12 (6.7)  & 9.4 (5.4) \\
Delayed SN     & 0.2  & 0.1  & 0.5  & 0.4  & 0.8 & 0.6 & 0.9 & 1.7 (0.9) & 1.3 (0.7) \\
High BH Kicks  & 0.04 & 0.03 & 0.09 & 0.07 & 0.1 & 0.1 & 0.3 & 0.6 (0.2) & 0.5 (0.2)\\
\hline
{BH-BH} \\
Standard       & 56  & 66  (61)  & 106 & 153 (140) & 183  & 246  (235)  & 610  & 369 (226)   & 514 (292) \\
Optimistic CE  & 287 & 324 (297) & 629 & 828 (745) & 1124 & 1421 (1339) & 3560 & 2384 (1336) & 3087 (1633) \\
Delayed SN     & 53  & 64  (59)  & 97  & 152 (139) & 171  & 241  (231)  & 596  & 345 (213)   & 501 (291) \\
High Kick      & 0.9 & 1.5 (1.4) & 1.4 & 3.8 (3.6) & 3.2  & 5.9  (5.8)  & 19   & 6.6 (4.0)   & 13 (7.2)
\enddata
\label{rates2genL}
\tablenotetext{a}{Same as Table~\ref{rates2genH}, but for the
  \textit{low-end} metallicity scenario.}
\end{deluxetable*}

\section{Results} \label{sec:results}

In Section~\ref{sec:simplerates}
we obtained a rough estimate of event rates by extrapolating the local
rate density via the scaling of 
Eq.~(\ref{eq:LocalUniverseMergerRateFormula}).
This extrapolation is expected to
provide a good approximation for low-mass systems (and in particular,
NS-NS binaries), because in this case the early inspiral makes up
most of the signal observable by advanced GW detectors, the signal extends through the detector band, and the detector range is sufficiently low that cosmological corrections to detectability and the dependence of merger rates on redshift can largely be ignored. The approximation will
become increasingly inaccurate for high-mass binaries, such as those
comprising one or two BHs. In Sections~\ref{sec:IMR} and~\ref{sec:fullrates}
we went beyond this
approximation by implementing three ``complete'' IMR waveform models
(EOB, PhC, PhB), and we described how to combine these models with
simulations from the {\tt StarTrack} code in order to obtain more
accurate estimates of the event rates (see Eq.~(\ref{eq:rate})).

The analytical estimates of Section~\ref{sec:simplerates} with local
merger rates based on the {\tt StarTrack} code are presented in
Table~\ref{tab:simplerates}. The more careful event rate calculations
of Section~\ref{sec:fullrates} are listed in Table \ref{rates2genH}
(for the high-end metallicity scenario) and Table \ref{rates2genL}
(for the low-end metallicity scenario).  

In these tables, the ``single-detector'' columns represent estimated detection rates
for a single detector with a $\rho \ge 8$ threshold for detectability. This is often used 
as a proxy for rates in multi-detector networks \citep{2010CQGra..27q3001A}.  In the 
``three-detector'' columns we consider two alternate detectability thresholds: minimum 
{\it network} SNRs of either 10 or 12 for a three-detector network composed of three instruments 
located at the LIGO Hanford, LIGO Livingston, and Virgo sites, all with aLIGO sensitivity. The 
network SNR threshold of 10 would have yielded false alarm rates of roughly once per decade in 
2009-2010 initial LIGO and Virgo data \citep[see Fig.~3 in][]{scenarios}.  This
threshold is optimistic for making confident detections if data quality in advanced detectors is 
similar to that in the initial detectors and the same searches are used.  With this in mind, 
\citet{scenarios} assumed a network SNR threshold of $12$ with an additional threshold constraint
on the SNR in the second-loudest instrument; we consider a simple SNR threshold of $12$.
Detection rates using a network SNR threshold were calculated using the same
framework as above, but implementing a network-geometry-dependent
$P(w)$ described (and fitted) in the Appendix.  
In the order-of-magnitude estimates described by
Eq. (\ref{eq:LocalUniverseMergerRateFormula}) and provided in Table
\ref{tab:simplerates} we employ $\left<w^{3}\right> \simeq 0.404$ for
the three-detector network ($\rho \ge 10$), a factor of $\sim
4.6$ larger than the value $\left<w^{3}\right> \simeq (1/2.26)^3
\approx 0.0866$ used for a one-detector network.

We now discuss these rate predictions, their dependence on
gravitational waveform models, and the astrophysical properties of DCO
populations observable by advanced GW detectors.

\subsection{Broad features of rate estimates}

The main conclusion of this work is
that BH-BH mergers should yield the highest detection rates in all
advanced detectors (aLIGO, AdV, and KAGRA), followed by NS-NS
mergers, with BH-NS mergers being the rarest. This finding is independent of 
our evolutionary models and of the details of the gravitational waveforms 
(however, see Sec.~\ref{sec:conclusions} for discussion).
The only exception is the ``Optimistic CE'' model, where detection rates for
BH-NS mergers dominate over NS-NS mergers (with BH-BH mergers still
dominating the detection rates). This model makes the
assumption that CE events with HG donors do not always end in
a premature merger, allowing more binaries to survive the CE and form
merging DCOs, and therefore increasing detection rates. As a result
the Optimistic CE model yields very large BH-BH rates, comparable to,
though a factor of a few below, existing upper limits on the BH-BH
binary mergers from initial LIGO/Virgo observations \citep[see,
  e.g.,][]{comparison,2012PhRvD..85h2002A,2013PhRvD..87b2002A}.

Our quantitative predictions for compact binary merger rates are consistent
with our previous papers in this series \citep{dominik,dominik2}. In
particular, we agree with the main conclusion of those papers: detectable 
BH-BH binaries can be formed over a broad range of metallicities, with a 
significant proportion forming in highly subsolar environments
(Fig.~\ref{fig:FiducialResultDistributions:BHBH}).
On a model-by-model basis our results are in good
agreement with prior work, with factor-of-two or smaller differences due to 
our inclusion of cosmological effects.

As expected, the simple approximation of
Eq.~(\ref{eq:LocalUniverseMergerRateFormula}) gives a good
order-of-magnitude estimate of the NS-NS detection rates listed in
Tables \ref{rates2genH} and \ref{rates2genL}. However, the
approximation fails for BH-BH systems. By comparing the detection
rates from Table \ref{tab:simplerates} with inspiral rates from Tables
\ref{rates2genH} and \ref{rates2genL}, we see that the local Universe
approximation of Eq.~(\ref{eq:LocalUniverseMergerRateFormula})
overestimates more careful calculations of detection rates by a factor
$\sim 2$ for BH-BH systems.  The limited signal bandwidth of high-mass
systems, the redshift dependence of binary merger rates, and
cosmological corrections make simple scaling relations inaccurate over
the large volume in which detectors are sensitive to BH-BH systems.
On the other hand, as the merger--ringdown phase of these binaries
falls within the sensitive band of second-generation interferometers,
it provides a significant contribution to the SNR. Indeed, as can be
seen in Tables \ref{rates2genH} and \ref{rates2genL}, the full IMR
calculations increase the detection rates considerably.  However,
BH-BH detection rates computed with appropriate cosmological
corrections are still lower than local merger rates converted into
detection rates via the basic scaling of
Eq.~(\ref{eq:LocalUniverseMergerRateFormula}).

\begin{figure}
\includegraphics[width=0.86\columnwidth]{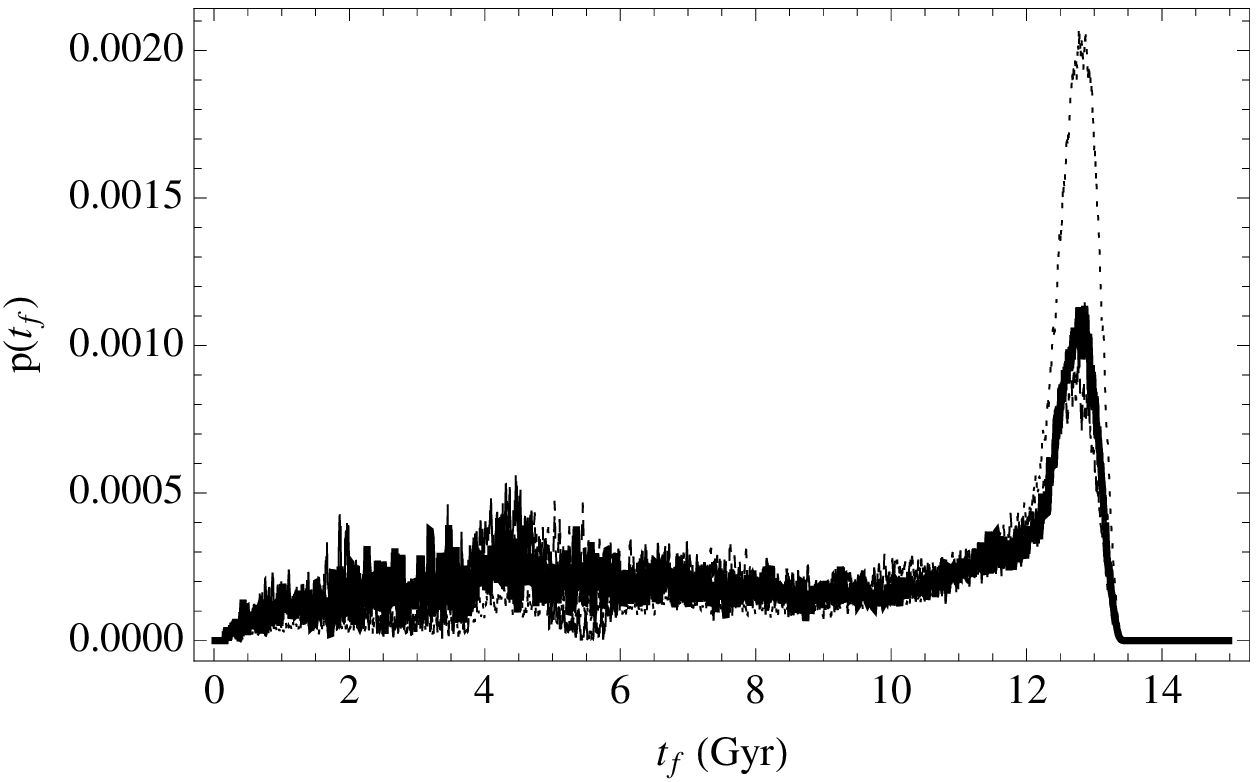}
\includegraphics[width=0.86\columnwidth]{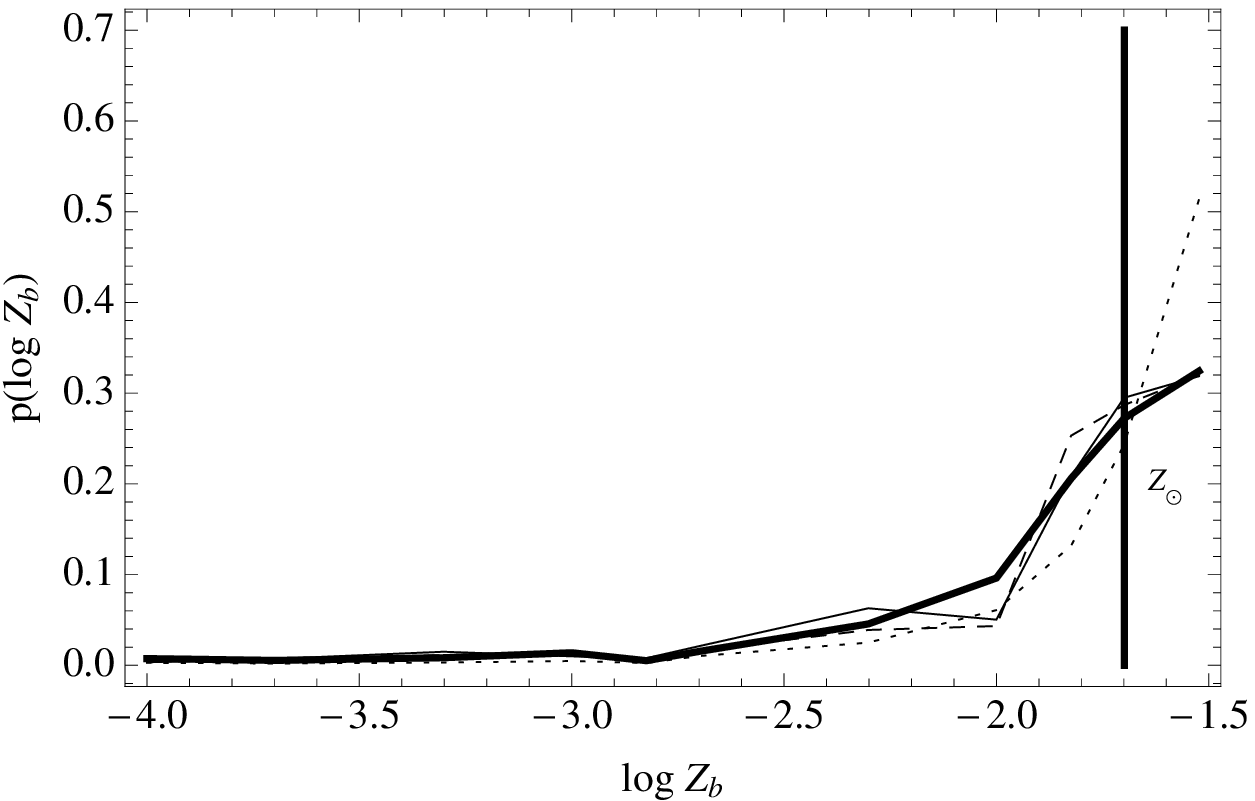}
\includegraphics[width=0.86\columnwidth]{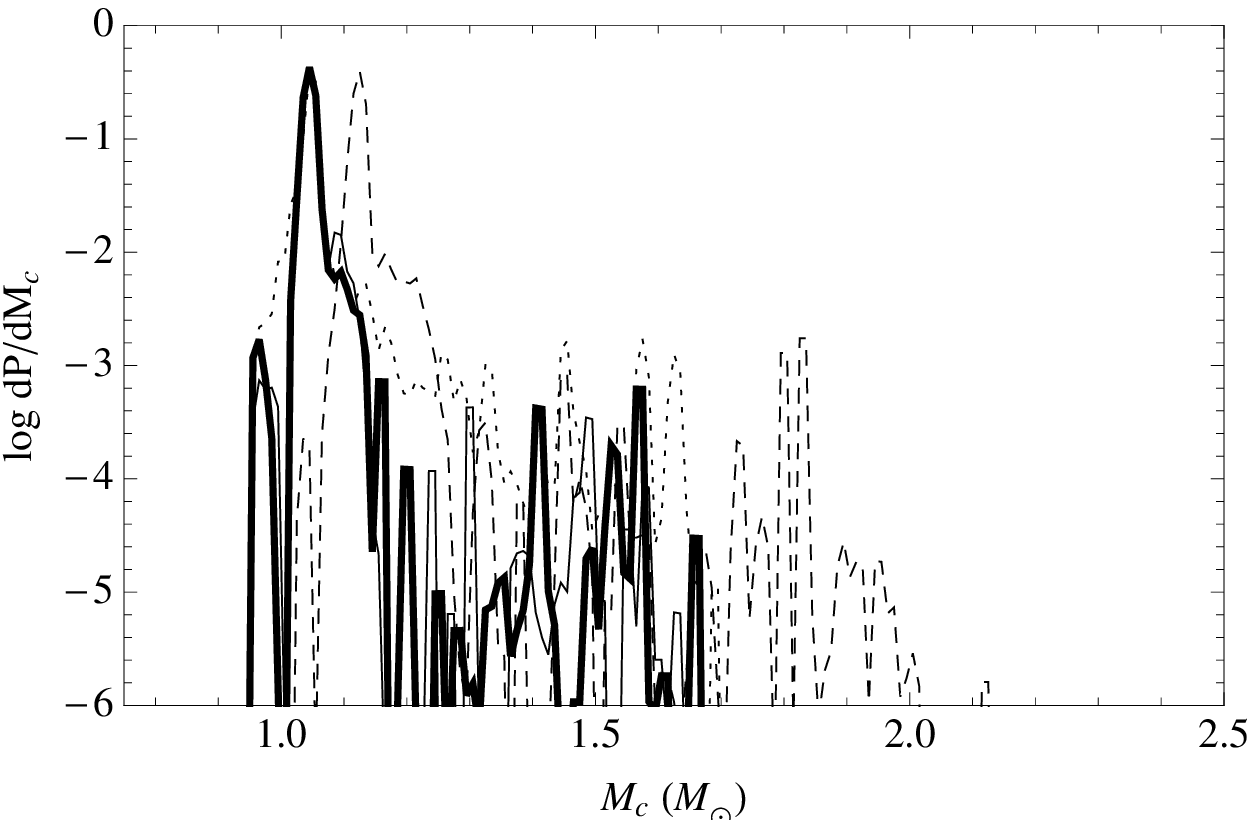}
\includegraphics[width=0.86\columnwidth]{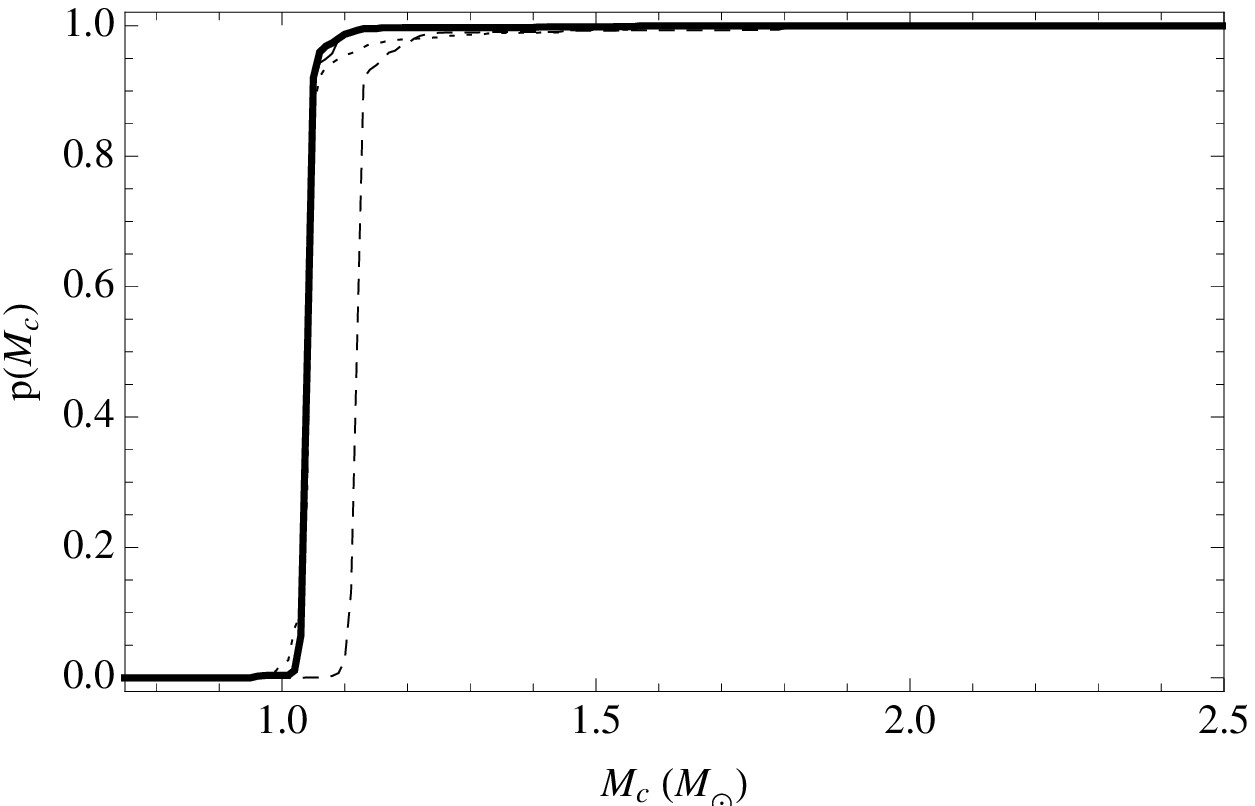}
\caption{ \label{fig:FiducialResultDistributions:NSNS}\textbf{Compact
    NS-NS binaries detectable by aLIGO}: Properties of NS-NS binaries
  with $\rho \ge 8$ in a single aLIGO instrument in the high-end metallicity
  scenario, scaled in proportion to their detection probability.
  Different color and line styles indicate results for different
  binary evolution models: Standard model (solid black), Optimistic CE (dotted black), delayed SN
  (dashed black), and high BH kicks (blue).  The top, second, and third panels show the
  distribution of birth time $t_{\rm f}$, birth metallicity $Z_{\rm b}$ (with a vertical bar
  marking solar metallicity, $\zsun=0.02$), and chirp mass
    $\mc$, respectively.  The bottom panel shows the cumulative distribution in chirp mass,
  to highlight significant changes on a linear scale.  The time domain ranges from $0$ Gyr (Big
  Bang) to $13.47$ Gyr (today). Though our simulations use a discrete
  array of metallicity bins, to guide the eye their relative
 contributions have been joined by solid lines in the second panel; this histogram makes
  no correction for the density of metallicity bins.
  }
\end{figure}

\begin{figure}
\includegraphics[width=\columnwidth]{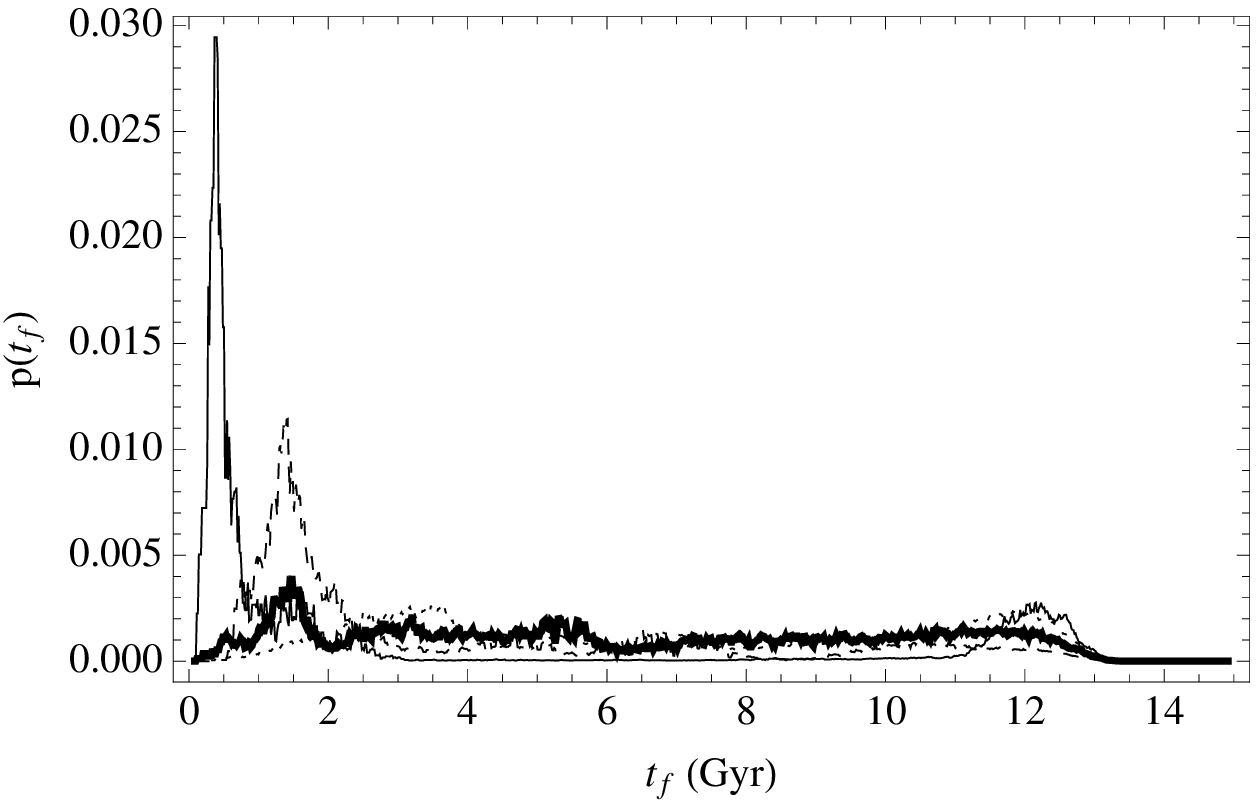}
\includegraphics[width=\columnwidth]{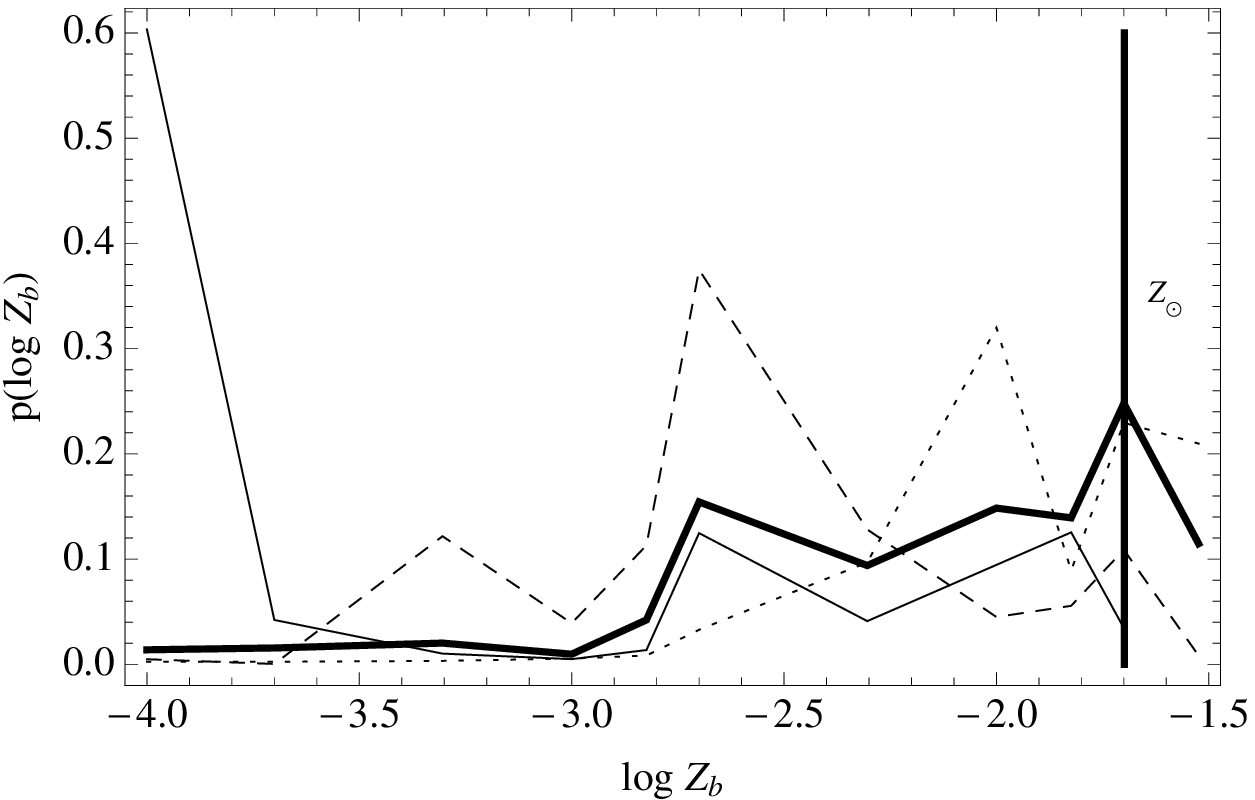}
\includegraphics[width=\columnwidth]{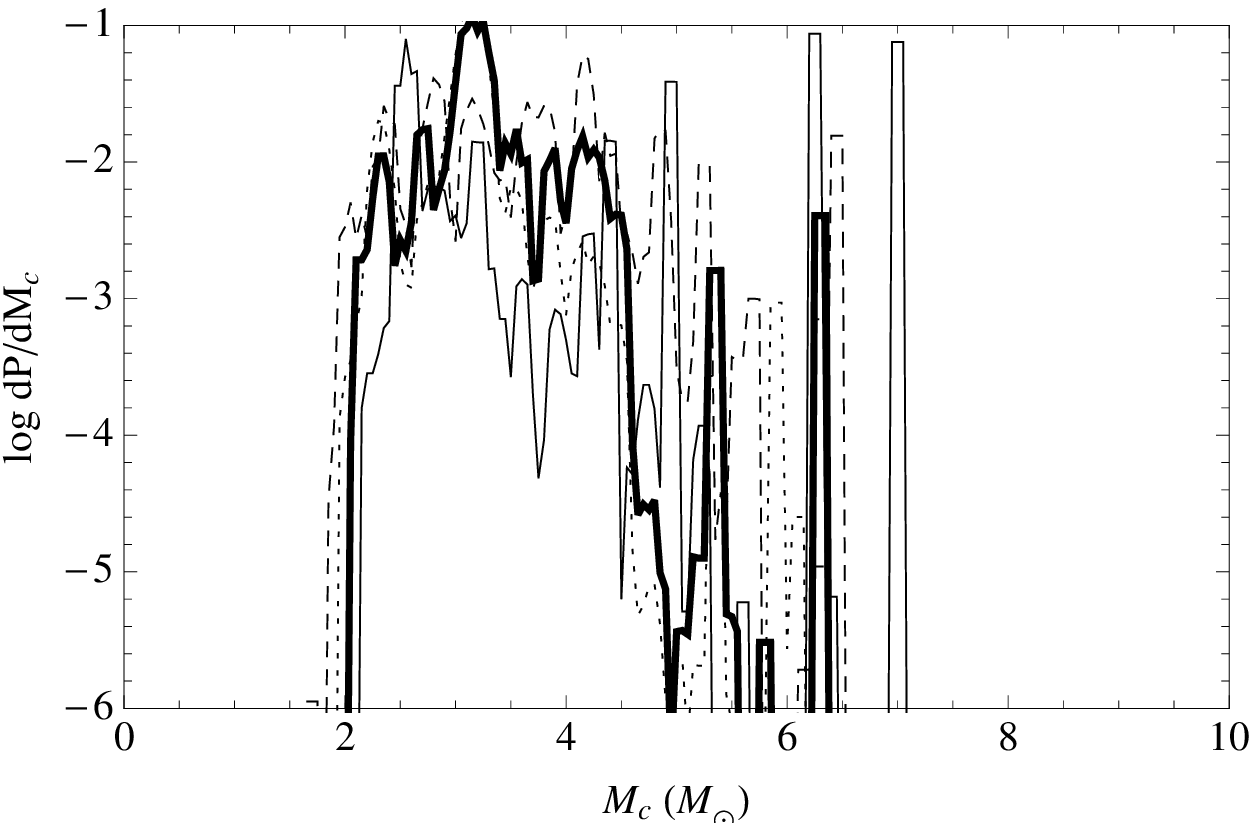}
\includegraphics[width=\columnwidth]{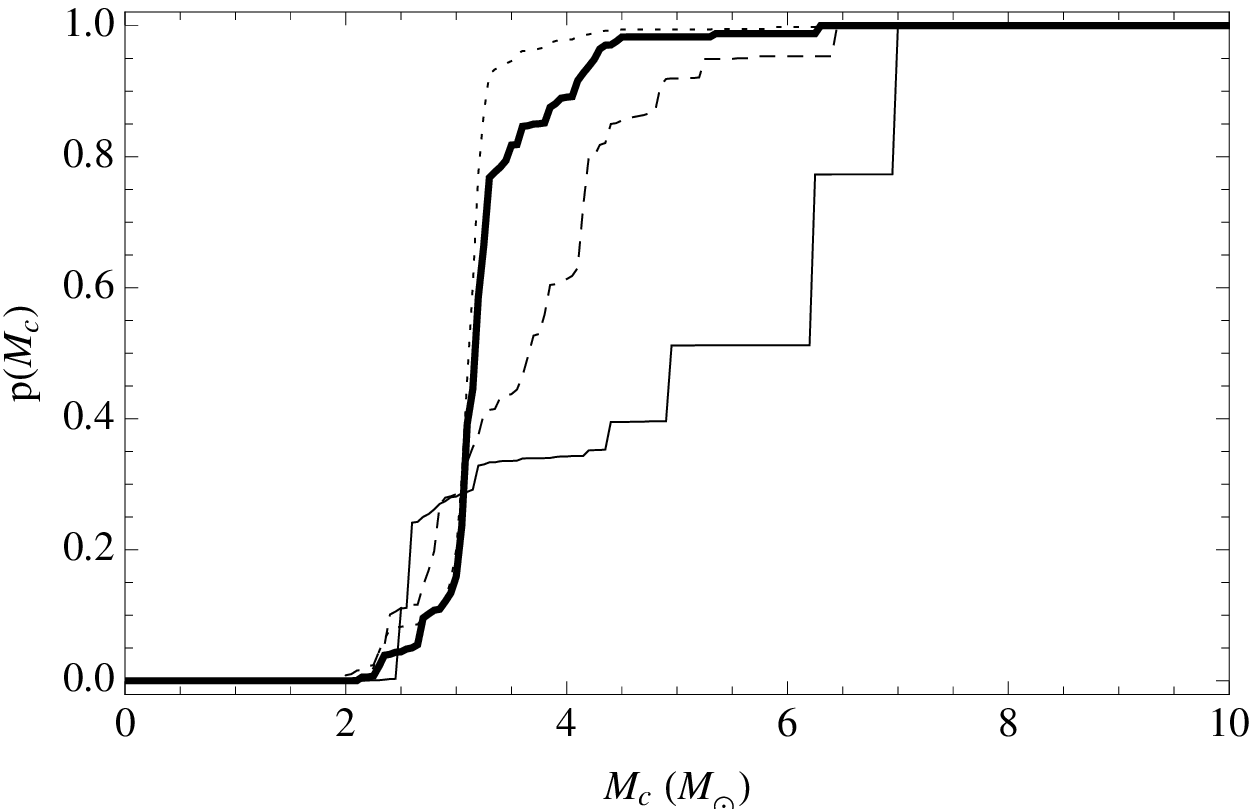}
\caption{\label{fig:FiducialResultDistributions:BHNS} \textbf{Compact
    BH-NS binaries detectable by aLIGO}: Same as
  Figure~\ref{fig:FiducialResultDistributions:NSNS}, but for BH-NS
 binaries in the high-end metallicity scenario.  Some of the sharp features in
 the chirp mass distribution are an artifact of the crude binning in metallicity
 undertaken for computational reasons; see the discussion in section \ref{dcos}.}
\end{figure}

\begin{figure}
\includegraphics[width=\columnwidth]{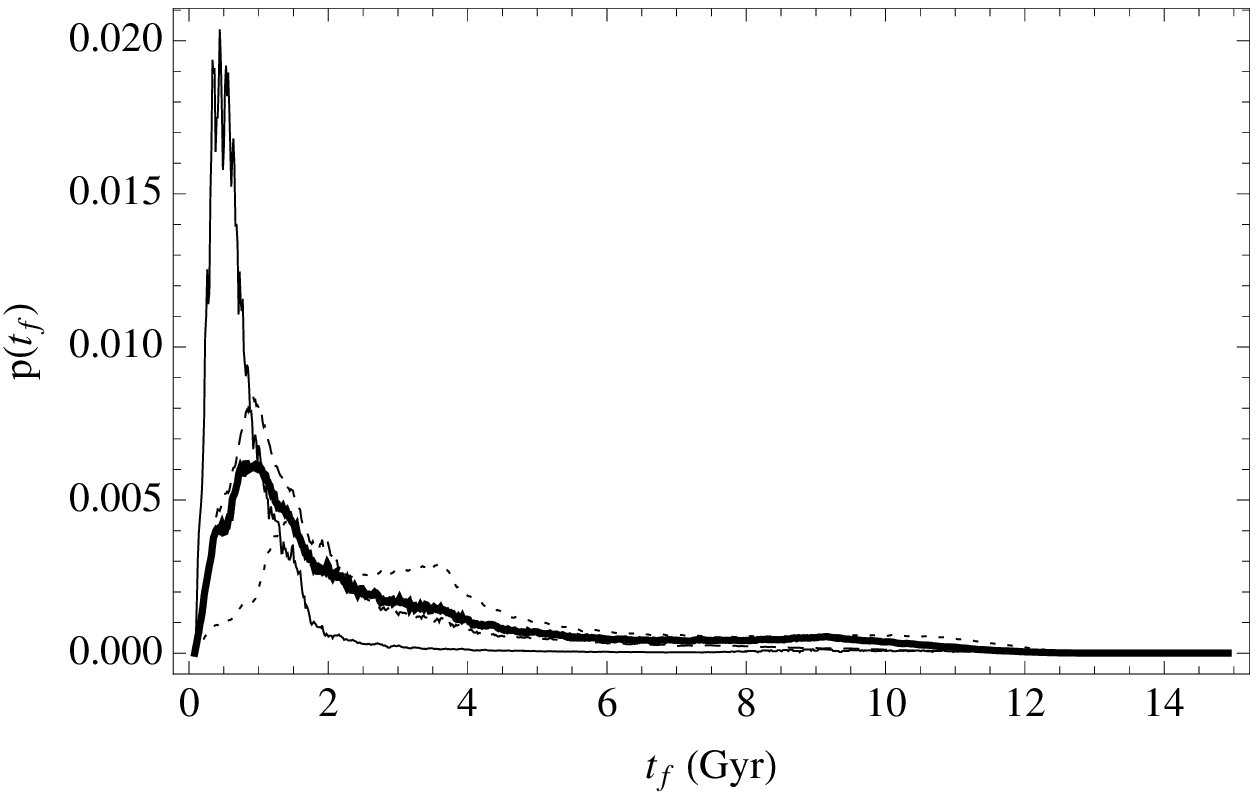}
\includegraphics[width=\columnwidth]{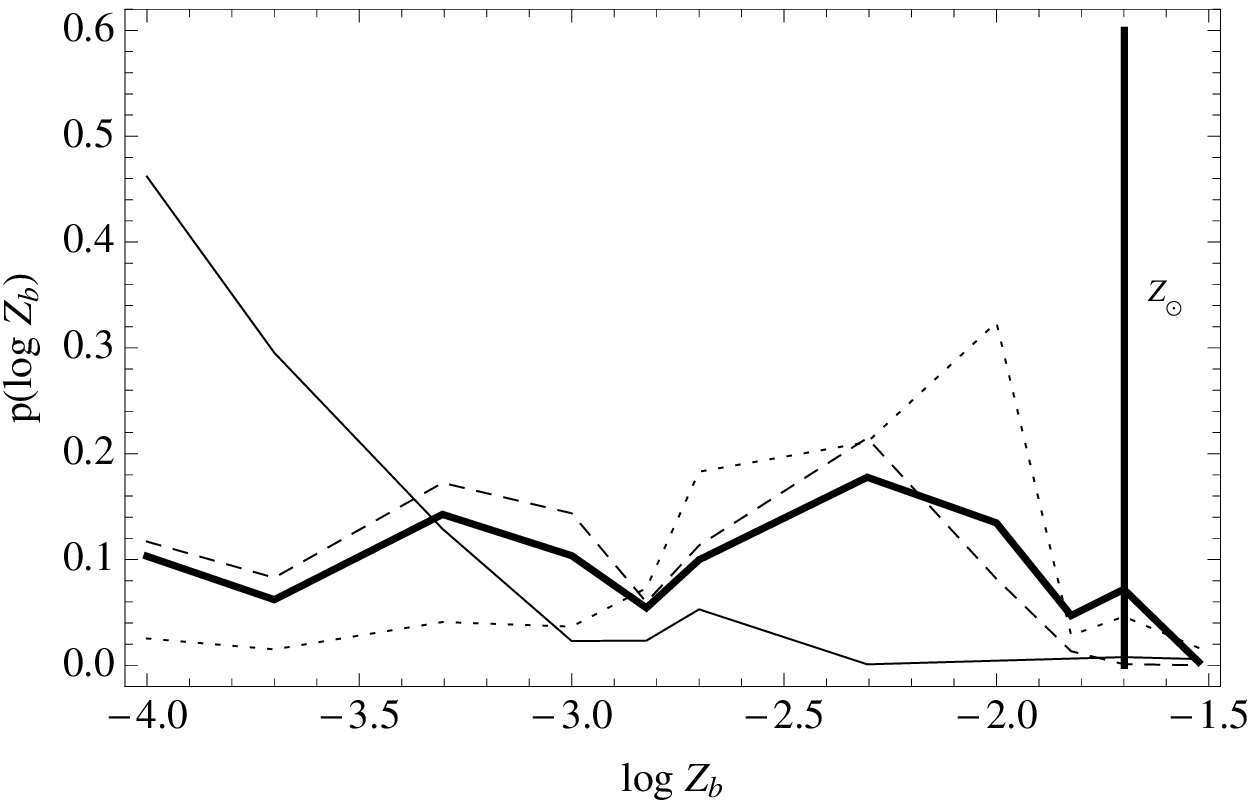}
\includegraphics[width=\columnwidth]{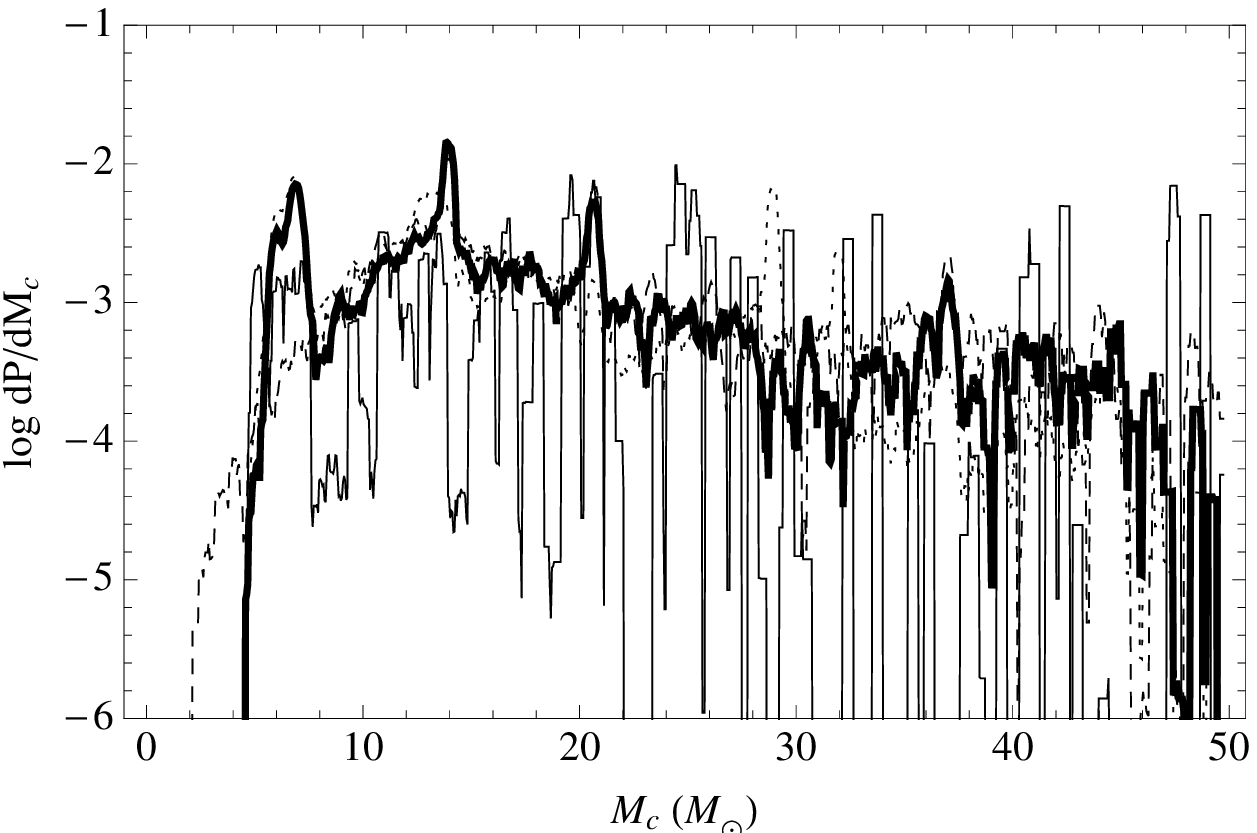}
\includegraphics[width=\columnwidth]{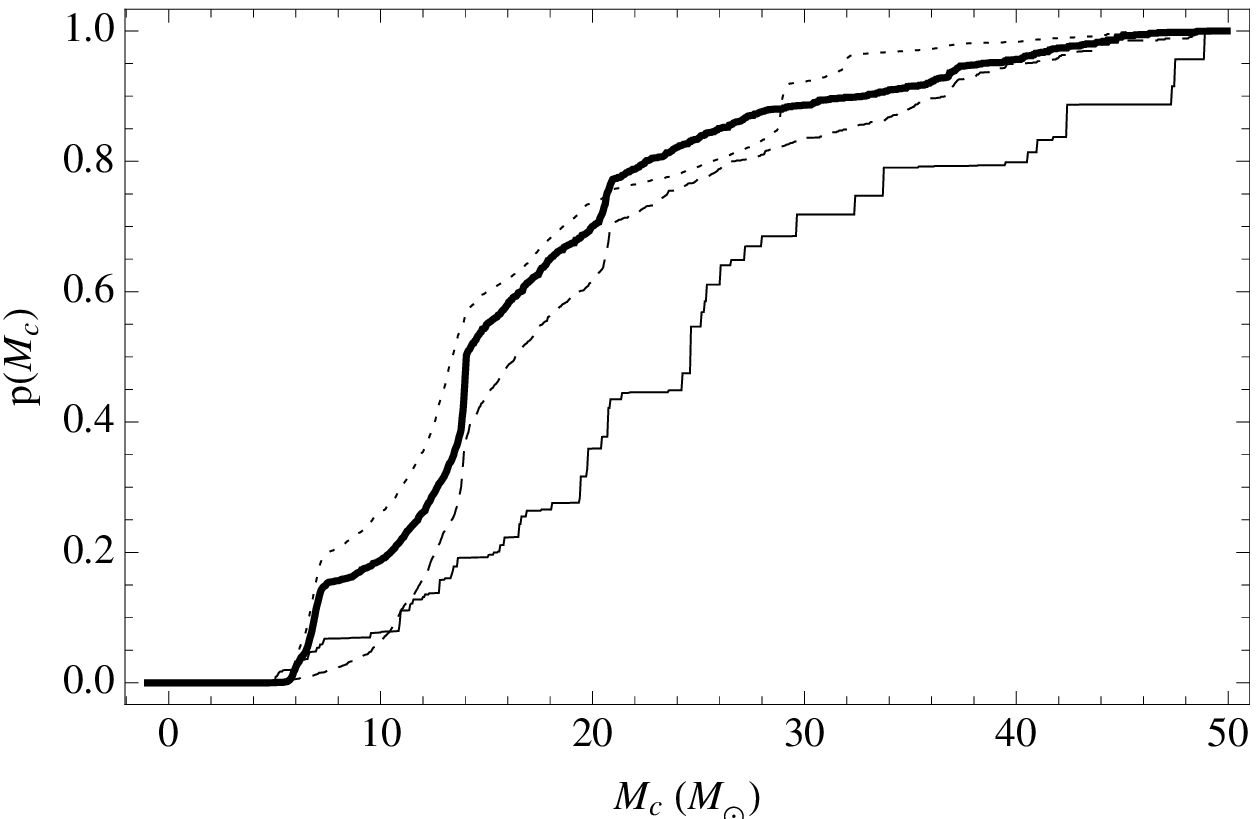}
\caption{\label{fig:FiducialResultDistributions:BHBH} \textbf{BH-BH
    binaries detectable by aLIGO}: Same as
  Figure~\ref{fig:FiducialResultDistributions:NSNS}, but for BH-BH
  binaries in the high-end metallicity scenario.  Some of the sharp features in
  the chirp mass distribution are an artifact of the crude binning in
  metallicity undertaken for computational reasons; see the discussion in section \ref{dcos}.}
\end{figure}

\subsection{Impact of waveform models on predicted rates}\label{subsec:WF}

Our results show that the merger-ringdown contribution is not
important for estimating detection rates of DCOs containing NSs. In
fact, when compared with the restricted PN model, the IMR waveforms
slightly {\em decrease} event rates for NS-NS and BH-NS systems.
The reason for this reduction is that IMR waveforms (such as PhC and
EOB) provide a more accurate representation of the early inspiral,
incorporating PN amplitude corrections that {\em reduce} the signal 
  amplitude\footnote{Note that in 
  Eq.~(3.14) of \cite{santamaria} the coefficient of the dominant correction, ${\cal A}_2$,
  listed in their Eq.~(A5) is negative.}---and hence the event rates---for
  signals dominated by the early inspiral. 

BH-NS systems may be subject to an additional event rate 
reduction mechanism. There is the possibility of the NS being 
distorted and disrupted by the BH tidal field. When these violent phenomena 
occur, a suppression of the GW amplitude takes place before the ISCO frequency,
and the SNR decreases with respect to that of a BH-BH system with the same 
properties. The GW shut--off due to NS tidal disruption depends on the parameters 
of the system: large values of the mass ratio, the BH spin, the NS
radius and the low tilt angles of NS orbital angular momentum relative to 
BH spin all favor NS disruption (e.g., \cite{kabelspin}).
By using point-particle IMR waveforms to describe the GW 
emission of BH-NS systems we are neglecting this event rate reduction mechanism. 
While it would be possible to take these effects into account for nonspinning systems by 
using the GW amplitude model of \cite{pannarale}, accurate models for systems
with spinning BHs do not exist yet. For consistency we therefore use 
BH-BH waveform models in both cases. Additionally, \cite{pannarale} found
that in the nonspinning case, the SNR difference between the mergers of disrupted 
BH-NS systems and the undisrupted systems modeled with PhC is less than $1\%$. 

Including the merger portion of the signal is important for BH-BH
systems. For illustration, let us focus on the Standard Model: if we
use PhC waveforms rather than the restricted PN approximation, we find a
  $\sim25\%$ increase in the detection rates of BH-BH
systems, from 117 (183) to 148 (246) in the
\textit{high-end} (\textit{low-end}) metallicity scenario. 

The rates predicted by EOB and PhC models agree quite well\footnote{We
  also carried out calculations using PhB models, which overestimate
  rates by about $10\%$ with respect to PhC models. We decided not to
  report these results in the Tables, because the PhB model is
  less accurate than PhC, although it is easier to implement and less 
  computationally expensive.}. This can be understood by looking again
at Figure \ref{fig:Ingredients:SNRVersusMass:CompareModels}, which
shows that different approximations of the strong-field merger
waveform agree rather well (at least in the equal-mass limit) on the
SNR $\rho$ and hence on the predicted event rates, which scale with
the cube of the SNR. Waveform differences produce systematic rate
uncertainties significantly less than a factor of 2, much smaller than
astrophysical differences between our preferred models.

Our detailed calculation shows that typically PhC models overestimate
the rates by about $10\%$ when compared to EOB models. This agreement
is nontrivial, because the two families of models are very different
in spirit and construction: the PhC family is a frequency-domain model
that can be easily implemented in rate calculations, while the
time-domain EOB model is more accurate in its domain of validity and
more computationally demanding. It is important to note that in order
to use the two families of models in rate calculations we must compute
waveforms and SNRs in regions of the parameter space where the models
were not tuned to numerical relativity simulations. In particular,
both models become less accurate for small mass-ratio binaries.

Besides systematic errors in waveform modeling, the detection rates
reported in this work (and the resulting distribution of detectable
DCO parameters) depend on our detection criteria.
We ignore a variety of complications of the detection pipelines, such
as the difficulty of searching for precessing sources, noise artifacts
(non-stationary, non-Gaussian ``glitches'' in the instruments) which
can make searches for shorter, high-mass signals less sensitive, and
the limited uptime of detectors. Instead, we have assumed several simplistic
detection thresholds on single-detector or network SNR that are constant across all masses and mass ratios.

Moreover, achieving good detector sensitivity at low frequencies may
prove particularly difficult.  We have only included bandwidth above
specified low-frequency cutoffs ($f_{\rm cut}=20$~Hz in most cases)
for detection-rate calculations.  However, the specific choice of low
frequency cutoff has minimal impact on our results. For example, using
a lower cutoff $f_{\rm cut}=10$~Hz rather than $f_{\rm cut}=20$~Hz in
the single-detector, high-end metallicity aLIGO rate calculation would
increase the Standard Model BH-BH rates from 117 to 128 in the
inspiral case, and from 148 to 161 in the IMR case. The effect is even
smaller for BH-NS and NS-NS rates.
  
The impact of spins on the predicted detection rates can be important.
We only consider BH spins, since NSs in compact binaries are not
expected to be rapidly spinning
\cite[e.g.,][]{MandelOShaughnessy:2010} and the dynamical impact of NS
spin will be small. In Tables \ref{rates2genH} and \ref{rates2genL} we
use the PhC model to estimate the possible impact of BH spin on BH-NS
and BH-BH detection rates by assuming that all BHs are nearly
maximally spinning (i.e., with dimensionless spin parameter
$\chi_1=\chi_2=0.998$) and aligned with the orbital angular momentum.
Aligned BH spins cause an orbital hang-up effect that increases the
overall power radiated in the merger, produces a rapidly spinning
merger remnant, and therefore increases the range to which high-mass
binaries can be detected.

We find that spin effects may increase BH-BH detection rates by as
much as a factor of $3$.  These increased rates are a direct result of
the increased horizon distance to spinning binaries. For example, a
(30+30) $M_\odot$ binary can be observed to roughly $1.3$ times
farther and be detected $\simeq (1.3)^3 \simeq 2$ more often with
near-maximal spins than with zero spin. Additionally, spin dynamics
can provide a direct diagnostic of the dominant physical effects in
DCO formation \citep{gerosa}. Spin effects only marginally increase
BH-NS rates, but (as discussed at the beginning of this section) tidal
disruption, which we neglected, may have the opposite effect.

\subsection{Astrophysical properties of observable DCOs} 
\label{dcos}

We now turn to a more detailed analysis of the observable
properties of DCOs. For concreteness we will focus on aLIGO results
for the ``Standard model'' and nonspinning PhC waveforms, unless
stated otherwise.

\smallskip
\noindent
\textbf{NS-NS}.
By comparing Tables \ref{rates2genH} and \ref{rates2genL} we see that
the detection rates of NS-NS systems are not sensitive to our
differing metallicity evolution scenarios. For simplicity, we therefore only
discuss our results for the \textit{high-end} metallicity evolution scenario.

As shown in our previous work \citep{dominik}, NS-NS systems are
efficiently created in metal-rich environments. The observable
population shares this trait, and half of the observable systems
originate from solar metallicities and higher. As the average
metallicity content of the Universe correlates with time and as most DCOs 
preferentially merge shortly after formation
(i.e., the time delay distribution is $\propto t_{\rm merger}^{-1}$; see \cite{dominik}), 
the birth rate of detectable NS-NS systems peaks at $13$ Gyrs after the Big Bang
(see Fig.~\ref{fig:FiducialResultDistributions:NSNS}).  The most
distant detectable system has a merger redshift $z\sim 0.13$ (or luminosity distance
$L_{\rm D}=610$ Mpc).

The range of possible chirp masses in the third panel from the top of 
Figure \ref{fig:FiducialResultDistributions:NSNS} is limited at the low end ($>0.87 M_\odot$) 
by the $1M_\odot$ minimum birth mass for NS and is limited at the high 
end  by the (assumed) maximum mass for a NS  ($m_{NS} <2.5 M_\odot; \mc < 2.1 M_\odot$). The birth mass, 
in turn, is set by supernova physics, which we have implemented as the Rapid or Delayed SN engine 
\citep{chrisija}. For this reason the NS mass difference between the SN engines is intrinsic 
to the entire merging population of NS-NS systems. Therefore, this observable feature should be
available to any of the detectors considered in this study.

The chirp mass distributions for Standard and Optimistic CE models span the range from 
$0.9\msun$ to $1.6\msun$. The Delayed SN model results in a notably different NS mass 
distribution, favoring heavier masses. As the SN explosion in the Delayed engine lasts longer, 
more matter is accreted onto the proto--NS (which is more massive than in the Rapid engine scenario),
allowing the formation of more massive remnants (cf.~Figure~\ref{fig:FiducialResultDistributions:NSNS}).
The maximum allowed NS mass in this model is $2.5\msun$, and in extreme 
(but very rare) cases this mass is approached; the maximum chirp mass 
for a detectable system in our Monte Carlo simulation was $2.1\msun$,
corresponding to both components close to the maximum allowed limit.  For comparison, 
chirp masses  of NS-NS systems in the models utilizing the Rapid SN engine 
(Standard, Optimistic CE and High BH kick) never exceed $1.7\msun$.  
Such extremely high masses are rare for all engines, however, and the majority 
of chirp masses are much lower, as seen in Figure \ref{fig:FiducialResultDistributions:NSNS}. 
The presence of more massive systems in the Delayed SN models extends the horizon
of NS-NS detectability to $z\sim 0.16$ ($L_{\rm D}=765$ Mpc). 

Lastly, we note that Standard and High BH kick models are identical for NS-NS systems.  
The difference between the black curve (Standard) and blue curve (High BH Kick) in 
Figure \ref{fig:FiducialResultDistributions:NSNS} corresponds to the systematic errors 
associated with Monte Carlo errors of binary simulations, galaxy sampling, metallicity 
binning, etc.

\smallskip
\noindent
\textbf{BH-NS}.
In our previous study \citep{dominik2} we showed that BH-NS
systems are efficiently created at moderate metallicities (${\rm Z}
\sim 0.1\,\zsun$, or $\log({\rm Z})\sim -2.7$). Indeed,
Figure~\ref{fig:FiducialResultDistributions:BHNS} shows that about
half of all detectable BH-NS systems will originate from metallicities
${\rm Z}<0.5\,\zsun$ ($\log({\rm Z})< -2$).
These systems have higher chirp masses than NS-NS systems, on average
$3.3\msun$ vs. $1.2\msun$, and therefore the detectors can
sample BH-NS systems from a larger volume. However, BH-NS systems are
the rarest of all DCOs per unit (comoving) volume.  As a consequence,
BH-NS binaries typically yield the lowest detection rates. One
exception is the Optimistic CE model, in which the merger rate per
unit volume is large enough (while still being lower than for NS-NS
systems at all redshifts) that BH-NS detection rates are larger than
NS-NS rates because they are observed farther (cf. Table~\ref{tab:simplerates} and
Figure~\ref{fig:FiducialResultDistributions:BHNS}).

In our standard model BH-NS systems are detectable up to redshift
$z\approx 0.28$ ($L_{\rm D}=1.4$ Mpc).  However, in the Delayed SN model this value reaches
$z\approx 0.31$ ($L_{\rm D}=1.6$ Mpc). As discussed earlier, this is due to the more massive
NSs (up to $2.4\msun$) produced by the Delayed engine.

\smallskip
\noindent
\textbf{BH-BH}.
As discussed in our previous papers in this series
\citep{nasza,dominik,dominik2}, BH-BH systems are formed most efficiently in
low-metallicity environments. The detectable population reflects this
property: about half of all detectable BH-BH systems were created in
environments with metallicities ${\rm Z}<0.1\,\zsun$ ($\log({\rm Z})<
-2.7$).
As in prior studies \citep{nasza,dominik,dominik2,vosstauris}, our calculations
imply that BH-BH systems yield the highest detection rates for ground-based
interferometers. This is true even in the ``High BH kick'' model,
where the vast majority of binaries containing a BH are disrupted.

Adjusting the metallicity evolution in the Universe from
\textit{high-end} to \textit{low-end} we see a factor of $\sim 2$
increase in detection rates. In the \textit{low-end}
scenario the average metallicity in the Universe is lower at all times.
Low metallicity environments are much more effective at producing merging 
BH-BH systems than higher ones, hence the increase in the detection rates.

Half of the detectable objects have chirp masses above $14\msun$.   
The most massive of these systems originate from environments with very low metallicity 
content (${\rm Z}\sim 0.01\,\zsun$). The birth times of detectable BH-BH systems
peak at $\sim1$~Gyr after the Big Bang. Additionally, half of these systems 
were created within $\sim 2$ Gyrs of the Big Bang (see top panel of Figure
\ref{fig:FiducialResultDistributions:BHBH}), when the average
abundance of heavy elements was much smaller than today.  

As seen in Tables~\ref{rates2genH} and~\ref{rates2genL}, the detection rates of 
BH-BH systems vary as we change our assumptions between the four models and two metallicity
evolution scenarios. By comparing detection rates, for example, found by aLIGO with PhC waveforms, 
for the \textit{high-end} metallicity model (works for all model choices), we can distinguish two 
extreme configurations: (1) The High BH kick model yields the lowest rates of merging BH-BH systems 
($3.8$ yr$^{-1}$). This is a direct consequence of assuming the presence of the maximum natal kick 
velocities allowed within our framework, which efficiently disrupt BH progenitor binaries. (2) The 
highest detection rate is achieved with the Optimistic CE model ($618$ yr$^{-1}$). Here, it is assumed 
that binaries are allowed to progress through the CE with a HG
donor, which adds a significant amount of BH-BH systems to the detectable population. 
The detection rates of the other two models: Standard and Delayed SN are similar to each other
($148$ yr$^{-1}$ and $129$ yr$^{-1}$, respectively).

The farthest objects are detectable out to $z\sim2$ ($L_{\rm D}15$ Gpc). These 
systems consist of the most massive BH pairs ($m_1=61\msun$ and $m_2=66\msun$ in the detectable
population, with a chirp mass equal to $55\msun$), born $1.8$ Gyr after the Big Bang, 
and originating from regions with our lowest considered metallicity content (${\rm Z}=0.005\,\zsun$). 
Note that the maximum mass of BH-BH systems is limited by the maximum ZAMS mass of stars, 
which was set to $150\msun$ in the current simulations. The effect of IMF extending to much
higher masses on detection of BH-BH inspirals have been recently presented by \cite{walczak}.

The detectable BH-BH chirp mass distribution for the Standard model has three 
major peaks. These are present at $\sim 7\msun$, $14\msun$, and $21\msun$ 
(see the black curve in the 3rd and 4th panels of Fig.~\ref{fig:FiducialResultDistributions:BHBH}).
Their presence is associated with the physics governing the Rapid SN engine and the formation 
of the most massive BH-BH systems. Within this framework we can distinguish three scenarios 
for BH formation, each depending on the pre-SN carbon--oxygen (CO) mass (see Eq.~16 in \cite{chrisija}).
The ``\textit{A}'' scenario occurs for $6\msun < M_{\rm CO} \leq 7\msun$ and results in full fallback on the BH
and, therefore, no natal kicks (see Eq.~\ref{vkick}). The ``\textit{B}'' scenario occurs for
$7\msun < M_{\rm CO} \leq 11\msun$, where the fallback is partial and some natal kicks 
are present. For this scenario we expect a decreased number of BH-BH systems because of 
natal kicks disrupting binary systems during SNe. The ``\textit{C}'' scenario develops for 
$M_{\rm CO} \geq 11\msun$ and again results in full fallback, and no natal kicks.

BH progenitors originating from $\zsun$ environments never form through the \textit{C}
scenario, since they lose mass in winds at rates that do not allow them to form CO cores larger than 
$11\msun$. Since BH-BH progenitors in the \textit{B} scenario are subject to disruption due to
the presence of natal kicks, most BH-BH systems in $\zsun$ environments form through the \textit{A} 
scenario, with chirp masses clustered around $7\msun$. 

However, reducing the metallicity by a factor of $2$ lowers the wind mass loss rates
sufficiently to allow BHs to form through the \textit{C} scenario. At this metallicity ($\sim 0.5\,\zsun$)
only the most massive progenitors ($M_{\rm ZAMS}>100\msun$) may form BHs through
this scenario. Additionally, the mass of the BHs formed from these high mass components 
($M_{\rm ZAMS}>100\msun$) only depends weakly on their initial mass. This stems from the fact that 
these stars evolve quickly ($\sim\,\mbox{Myrs}$) and lose large fractions of their hydrogen envelope. 
Binary evolution does not alter this result significantly, as the interactions between components, 
such as mass transfer during CE episodes, also lead to the removal of their hydrogen envelopes.
The result for metallicity $\sim 0.5\,\zsun$ is a clustering of BH-BH systems formed 
from the most massive binaries at masses around $16\msun$ for each component.
This produces the peak in the chirp mass distribution at $\sim 14\msun$. 

Reducing the metallicity content by another factor 
of $2$ (to $\sim 0.25\,\zsun$) allows the same mechanism to form BH-BH systems with masses 
clustering at around $24\msun$ for each component. These systems form the peak in the chirp mass
distribution at $\sim 21\msun$. 

The grouping effect disappears when reducing the metallicity abundance in BH progenitors 
even further. For example, at $0.1\,\zsun$ the low wind mass loss rate does not increase the
separation between components as significantly as for higher metallicities. Consequently, the most 
massive progenitor binaries engage in a CE phase early in their evolution. This usually happens when 
the donor is on the HG and the Standard model does not allow for successful outcomes of such CEs. 
However, this scenario is allowed to form BH-BH systems in the Optimistic CE model, yielding the peak 
present in the chirp mass distribution at $\sim 29\msun$. 

As discussed above, the chirp mass distribution in scenario \textit{C} 
depends sensitively on the mass loss rate of stars, which depends strongly on metallicity. 
Binary evolution for $0.5\,\zsun$ and $0.25\,\zsun$ creates sharp peaks in the chirp mass
distribution of BH-BH systems. In the discrete metallicity grid simulated in this study, 
there are no metallicity points between $0.5\,\zsun$ and $0.25\,\zsun$. Targeted follow-up investigations 
indicate that metallicity choices between $0.5\zsun$ and $0.25\zsun$ lead to additional sharp peaks in 
the chirp mass distribution between $14\msun$ -- $21\msun$.  We expect that an integral over a 
fine grid with appropriately small step sizes in metallicity would lead 
to all of these narrow peaks merging together to form a single broad distribution without 
sharp features.  However, we cannot confidently describe the shape of this distribution 
without a more detailed investigation with a fine grid of metallicities, which is not 
computationally tractable at present. 

Finally, the peak in the chirp mass distribution at $\sim 7\msun$ in the Standard model is 
formed from systems born in $0.5$--$1\,\zsun$ environments. These are low-mass BHs (usually
$8$--$9\msun$ per component) formed in the \textit{A} scenario. This formation
is particularly interesting as it does not appear in the Delayed SN 
model, with the difference stemming from the different fallback scenarios in the Rapid and Delayed
engines. With the Rapid engine, we can distinguish the three fallback regions.
However, the Delayed engine predicts one region of partial fallback for $3.5\msun < M_{\rm CO} \leq 11\msun$ 
and one region of full fallback $M_{\rm CO} \geq 11\msun$ (identical to the \textit{C} scenario in the
Rapid engine). Since partial fallback implies the presence of natal kicks and, therefore, 
increased probability of binary disruption, there are no ``preferred'' masses for the lightest 
BHs in the Delayed SN engine (see dashed line on the 3rd panel, 
Fig.~\ref{fig:FiducialResultDistributions:BHBH}) as in the Rapid engine.  

The Standard and Delayed SN models also yield different lower mass
limits for BH remnants (see Section \ref{binevol}). For the ``Rapid
engine'' scenario the lowest-mass BH is $\sim 5\msun$, while for the
``Delayed engine'' scenario the lowest-mass BH is $\sim 2.5\msun$
(this is also the highest NS mass adopted in our {\tt StarTrack}
calculations). As a result, the detectable systems with the lowest
total mass have $\mc=4.8\msun$ and $\mc=2.4\msun$ in the
Rapid and Delayed engine scenarios, respectively.

\begin{figure}
\includegraphics[angle=270,width=\columnwidth]{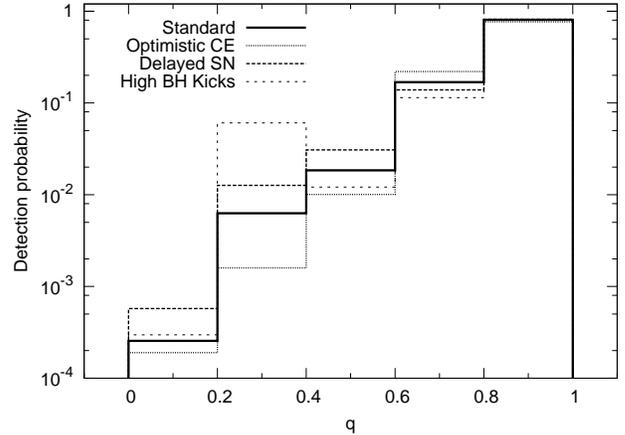}
\caption{\label{fig:qdis} \textbf{Mass ratio ($q$) detection probability distribution for BH-BH
systems.} It is clear that one should expect that the vast majority of detectable BH-BH systems
will be formed of nearly equal mass components. The lowest values of $q$ among the detected systems are
$0.05$ for the Delayed SN model and $0.12$ for the remaining models. For each model the probability
is normalized to the total number of detections for this model.}
\end{figure}

Additionally, regardless of our evolutionary models the majority BH-BH 
systems are formed with nearly equal mass components. Therefore, systems
with mass rations $\sim 1$ dominate the detected population, as shown
in Fig.~\ref{fig:qdis}. For the Delayed SN model the detectable BH-BH systems
with the lowest mass ratio have $q\approx 0.05$. For the remaining models
this value is $q\approx 0.12$. 

For future reference we also present the initial--final mass
relation for close BH-BH systems in Fig.~\ref{fig:bmr}. The relation is divided into the primary 
(more massive at ZAMS) and secondary (less massive) component for two metallicity values 
($\zsun$ and $0.1\zsun$), for the Standard model. It is clearly visible that 
binary evolution distorts the initial-final mass relation for single stars in both mass dimensions.
In the initial mass dimension, the absence of BHs forming from stars with ZAMS mass above $\sim 70\msun$ 
is a direct consequence of the assumption of the negative (merger) CE outcome for HG donors in our Standard 
model. In our framework more massive stars have larger radii and, therefore, are more likely to engage in 
CE while the donor is on the HG rather than on later evolutionary stages. If this assumption was relaxed 
(Optimistic CE model) the maximum BH mass reached in close BH-BH systems is found to be $150\msun$ for both 
metallicities. In the final mass dimension, binary evolution prevents remnant components
from reaching masses as high as those formed from single progenitors. Whereas single stars 
shed mass only through winds, binaries may also remove mass through interactions like the non-conservative 
mass transfer and/or CE events, which consequently lowers the mass of the remnants. 

\begin{widetext}
\begin{center}
\begin{figure}
\includegraphics[width=\textwidth]{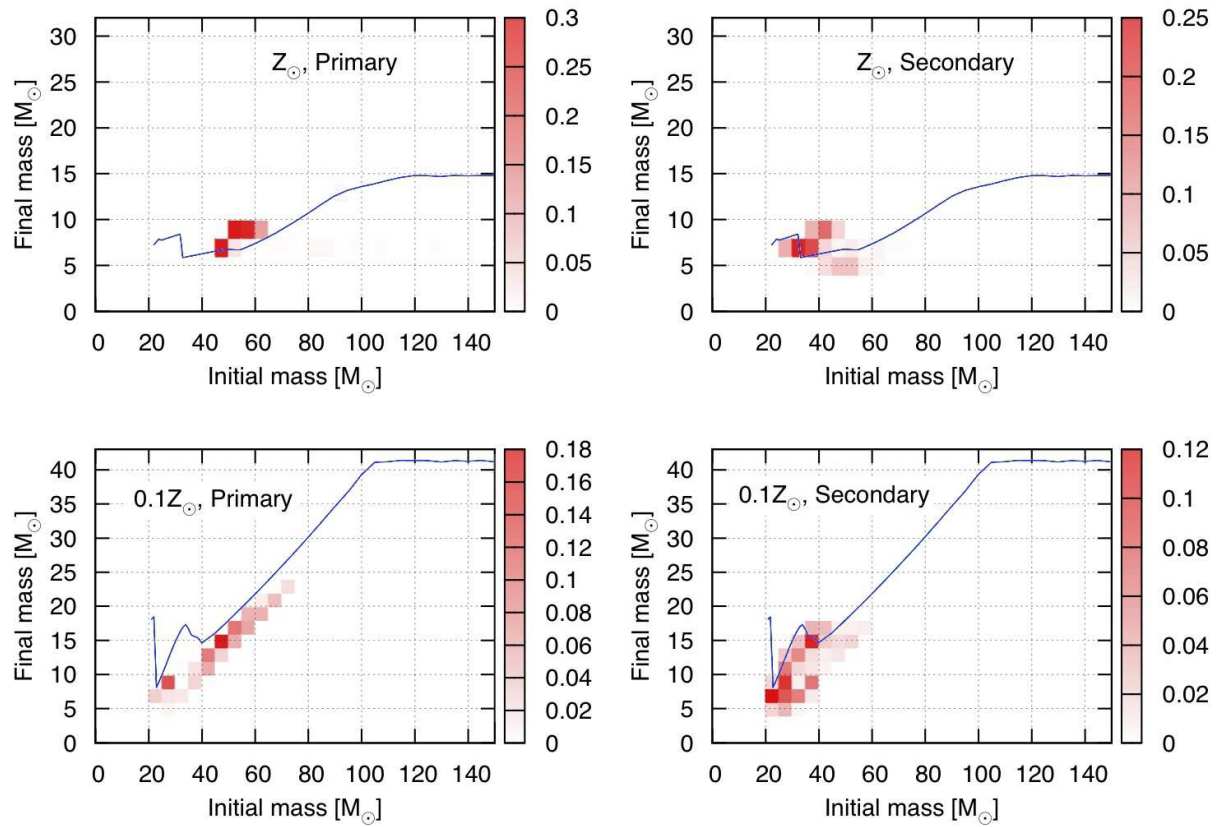}
\caption{\label{fig:bmr} \textbf{Initial-final mass relation for binary systems.} Presented
for close BH-BH systems, Standard model. We define primary and secondary components as the initially
(at ZAMS) more and less massive, respectively. The shaded scale (right side of each panel) shows the
fractional contribution of a given ZAMS mass bin to the total mass of merging black holes formed
from primaries (left panels) and secondaries (right panels). Note that binary evolution produces a very
different initial-final mass relation than the single stellar evolution (thin line). The top panels and
bottom panels show results for $\zsun$ and $0.1\zsun$, respectively.
}
\end{figure}
\end{center}
\end{widetext}

The initial-final mass relation (in this case for the binary population of
close BH-BH systems) is a result of a number of various initial and
evolutionary assumptions used in population synthesis calculations.  Change
of any of these assumptions (whether in initial conditions or evolutionary
calculations) may potentially influence the initial-final mass relation and in
turn the generated BH-BH population. The largest impact is expected from the
treatment of RLOF stability (i.e., criteria for CE development), SN
explosion physics, wind mass loss and internal mixing within massive stars
induced by convection and/or rotation that sets the radial evolution of massive
stars. It seems that the change in the assumptions underlying the initial-final 
mass relation may yield no BH-BHs \citep{mennekens} or numerous BH-BH systems 
\citep{voss,nasza,dominik,dominik2}. However, these results apply only to isolated binary evolution.
New studies of globular clusters suggest that, such environments may be the birthplaces of
a significant number of BH-BH systems \citep{gcbhbh}.

Note, that the above relations apply only to BH-BH systems. However, our models do not inhibit
the creation of NS from progenitors much more massive than $20 \msun$. In fact, the study by
\cite{betaam2008} shows that, due to binary evolution, NS may form from progenitors as massive
as $100 \msun$. 

\section{Questioning the no BH-BH theorem}\label{sec:nobhbh}
During more than a decade of research into the evolution of binary stars and the formation of 
DCOs, several authors proposed the absence of stellar-mass BH-BH systems merging within the Hubble time 
(e.g. \cite{nele2001,mennekens}). In the latter study the authors have claimed that the main 
reason for this are the high wind mass loss rates experienced by BH progenitors. For example, in 
their version of the Brussels population/galactic code (originally \cite{ddv04}) they fix the 
wind mass loss rates of the Luminous Blue Variable (LBV) phase at $10^{-3}\msun$ yr$^{-1}$. 
Following such heavy mass loss, the orbital separation of the components increases
so that they do not engage in CE. As the CE is a major mechanism for reducing orbital separation 
in isolated binary evolution, allowing for the formation of close BH-BH systems, the result is an 
absence of BH-BH systems detectable through gravitational waves. These results stand in contrast with the 
works of \cite{voss} and our previous studies \citep{nasza,dominik,dominik2}.

\begin{figure}
\includegraphics[angle=270,width=\columnwidth]{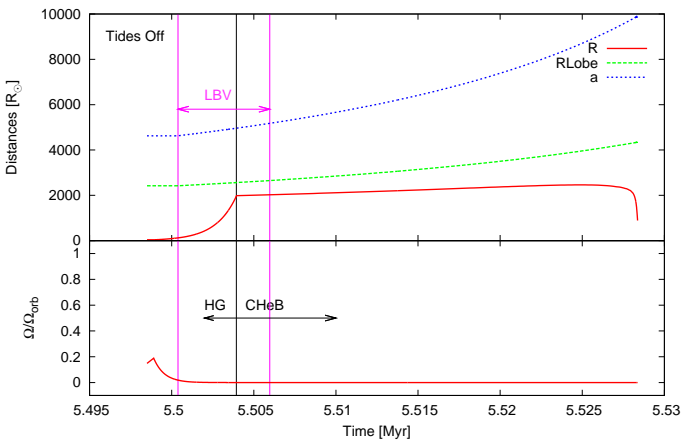}
\caption{\label{fig:notides} \textbf{Orbital evolution with tidal interactions disabled}.
This figure presents a part of the evolution of a  $8\msun$ BH and $43\msun$ HG system, with
Luminous Blue Variable wind mass loss rate set at $10^{-3}\msun$ yr$^{-1}$.
The top panel shows the evolution of the radius and Roche lobe of the HG star in
addition to the orbital separation in the binary. The bottom panel shows the evolution of the
HG star's spin frequency relative to the orbital frequency. The HG star's activity
as a Luminous Blue Variable is marked by the 'LBV' label. The vertical line separating the 'HG'
and 'CHeB' labels marks the transition of the HG star to the Core Helium Burning phase.
Note that without tidal interactions the binary's orbit expands (due to
stellar wind mass loss) and no component interaction (e.g., CE) is expected.
In the end a wide BH-BH binary is formed.
}
\end{figure}

There are mitigating factors to the finding of \cite{mennekens}. For example, their code does not allow
for tidal interactions between close binary components. As we demonstrate in the
following text, tidal interactions may (even for very high LBV winds) allow
for the formation of close BH-BH binaries (for more on the importance of tidal interactions see e.g., 
\cite{serena}). Let us consider the following example of binary
evolution generated with the StarTrack code. We start with an evolved binary: a $8\msun$ BH 
accompanied by a $43\msun$ companion at the beginning of the HG phase, with an orbital separation of 
$4600\rsun$ at $5.5$ Myr after the creation of the systems (ZAMS). This is a typical phase of a BH-BH progenitor
in our Standard model. In this example we also set the LBV wind mass loss rate to $10^{-3}\msun$ yr$^{-1}$
and disable tidal interactions between the components, both as in \cite{mennekens}. We find that
intense wind mass loss widens the orbital separation between the components to such extent that they
never interact. Therefore, when the BH companion forms a second BH, the resulting BH-BH
systems is too wide to merge within a Hubble time. This example is presented in Fig.~\ref{fig:notides}.

\begin{figure}
\includegraphics[angle=270,width=\columnwidth]{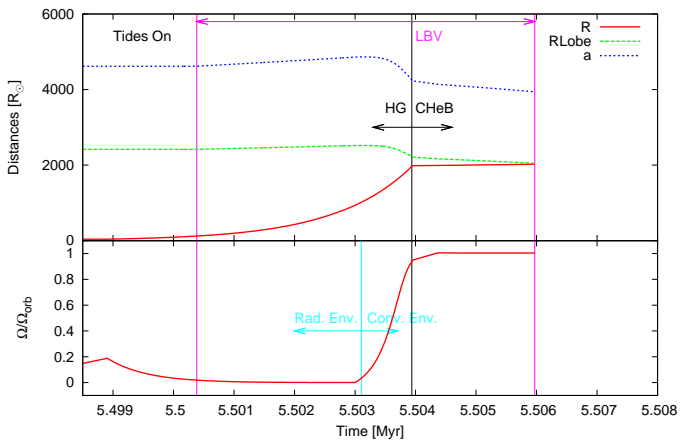}
\caption{\label{fig:tides} \textbf{Orbital evolution with tidal interactions enabled}.
Same as Fig.~\ref{fig:notides} but with tidal interactions enabled. The `Rad.~Env.' and
`Conv.~Env.' labels along with corresponding arrows highlight areas where the HG star has a radiative
and convective envelope, respectively. The vertical line linking the arrows marks the transition
point in the structure of the envelope.
Tidal interactions allow the transfer of orbital angular momentum into the
expanding HG star. The associated orbital decay leads to RLOF and the development of a
CE, which allows for the formation of a close BH-BH binary. The timescale on the horizontal axis is
zoomed in relative to Fig.~\ref{fig:notides}.
}
\end{figure}

We can repeat our exercise can be repeated with tidal interactions between the components enabled.
Investigating the same system we find a drastically different outcome of the evolution (see 
Fig.~\ref{fig:tides}). As in the example above, the BH companion starts its significant
evolutionary expansion across Hertzsprung gap. Due to the conservation of angular momentum, the expansion
of the star slows its rotation down almost to a standstill. 

Once the companion star fills a sizable fraction of its Roche lobe ($\sim 50\%$), the tidal torques 
imposed on the star by an orbiting BH transfer the orbital angular momentum into the star, spinning it 
up. At first this effect is negligible. However, after approximately $5000$ years, when the radius of 
the star becomes sufficient ($\sim 1100\rsun$), the spin up of the HG star stalls and overpowers the increase of 
orbital separation. From this point on, the orbital separation starts to decrease for another 
$3000$ years. Finally, when the radius of the star is $\sim 2000\rsun$, it fills its Roche lobe
and initiates a CE.

Our exercise clearly shows that different assumptions may lead to qualitatively different 
outcomes in terms of the close BH-BH formation. In particular, assumptions used in this study on 
LBV winds, tidal interactions and radial expansion result in a large number of BH-BH mergers.
In contrast, assumptions used by \cite{mennekens} result in no BH-BH mergers formed
out of the isolated binary evolution. 

There are several caveats in this framework. First, it is not theoretically well established if stellar
radii can grow to $\sim 2000\rsun$. For example, intensive mixing (either invoked by rapid rotation or
extended convection in the stellar interior) may reduce the size of the H-rich envelope
which is responsible for expansion in massive stars. On the other hand the
intense wind mass loss may additionally reduce the envelope (e.g., \cite{yusof2013}, 
but see MESA models for very massive stars \citep{walczak}).
However, the radii of AH Sco, KW Sgr and UY Scuti estimated with
the PHOENIX stellar atmosphere model \citep{wittkowski} extend well beyond $1000\rsun$, with UY Scuti, 
reaching $1708\rsun$ \citep{yuscuti}. The mass of UY Scuti is estimated to be within
$25\msun$--$40\msun$, i.e., within the mass range for BH progenitors in our framework. Second, the 
efficiency of tidal interactions depends on the structure of the envelope of the participating components. 
Stars with convective envelopes tend to respond more strongly to tidal dissipation than stars with radiative 
envelopes. In {\tt StarTrack} (see Section 3.3 of \cite{startrack}) we calibrate this phenomenon against 
the cutoff period for circularization of a population of MS binaries in M67 and the orbital decay 
accompanying tidal synchronization in the LMC X-4 high mass X-ray binary. 

This treatment of tidal dissipation applies directly to the given example as the envelope of the 
companion star turns from radiative to convective about $3000$ years after the companion enters the 
HG (when HG star radius increases to over $\sim 1000 \rsun$). 
However, our simulations show that switching 
tidal dissipation to the weaker radiative damping does not prevent binaries from 
initiating the CE. In our framework tides are applied to the entire star and 
we assume that stars rotate non-differentially. It cannot be excluded that  
tides operate only on the outer layers of stellar atmosphere that holds
only a small fraction of a star's mass. Additionally, if there is no (or very  
weak) transport of angular momentum within a star, only a
small fraction of orbital energy is used to synchronize the stellar atmosphere 
as compared to our prescription. Finally, the moment of inertia of very massive stars depends 
strongly on the radial profile, and the {\tt StarTrack} assumptions may yield a moment of inertia 
that is too large, therefore providing a more significant reservoir for depositing orbital angular 
momentum into the star than is available in practice. If in fact only very little orbital angular
momentum is used for binary component synchronization {\em and} if the winds are   
in fact as intense as indicated by \cite{mennekens}, then this would bar the 
formation of many close BH-BH binaries found within the framework of our 
evolutionary model.

Even if tidal interactions turn out to be ineffective in massive close binaries, this 
does not necessarily rule out the formation of close BH-BH binaries.  In field populations 
about 10--30\% of binaries are, in fact, triples (or higher multiples; e.g., \cite{kiminki1,kiminki2,duchene}) and 
Kozai-Lidov effects or dynamical instabilities \citep{PeretsKratter:2012} may lead to the merger 
of wide BH-BH binaries.
Additionally, many \citep{kroupa2014} massive stars are formed in clusters and may be subject to
dynamical interactions that can potentially decrease orbital separations.
Finally, over the last few years it has been claimed that dense globular clusters may produce significant 
number of close BH-BH binaries. In contrast with earlier findings with no efficient formation of close BH-BH
binaries (e.g., \cite{kulkarni,sigurdsson,zwart2000,banerjee}) the new paradigm  emerged based on 
recent and updated Monte Carlo simulations of dense cluster evolution (e.g., \cite{mackey,morscher,sippel,
heggie2014}). BH-BH binaries may also form via dynamical interactions in galactic nuclear clusters with or 
without a massive black hole \citep{OLeary:2008,MillerLauburg:2008} (but cf.~\cite{Tsang:2013}).

\section{Conclusions}\label{sec:conclusions}

We have calculated cosmological detection rates of merging DCOs for
second-generation GW observatories. We used redshift distributions of
merging DCOs from the {\tt Startrack} population synthesis code, and
have incorporated the cosmic star formation rate as well as galaxy and
metallicity evolution. Using state-of-the-art gravitational waveforms
and detector sensitivity curves, we have translated the cosmological
merger rates into detection rates for four distinct models of binary
evolution.

Our study has several robust implications for imminent GW
searches. First and foremost, our four models agree on the detection rates
of merging NS-NS systems ($\sim 1$ detection per year), with the exception of the Optimistic CE
model which predicts rates a factor of $2$--$3$ times higher than
other models. The mass distributions of detectable NS-NS systems are also similar across the models,
with the exception of the Delayed SN model, which allows for the
formation of NSs with higher masses due to prolonged accretion during
the SN explosion. We predict that NS-NS binaries will be detectable up to redshift
$z\approx 0.13$, i.e., only in the local Universe.
 
Second, BH-NS systems are expected to be the rarest detectable DCOs (less than $1$ detection
per year), with the exception of the Optimistic CE model, in which BH-NS
detection rates slightly exceed those of NS-NS systems of the same model. We predict
BH-NS systems to be detectable up to redshift $z\approx 0.3$.

In contrast, BH-BH systems will provide the largest number of
detections ($\sim 100$--$1000$ per year), making them the primary target for first detection 
and the most promising source for future statistical studies of source
populations. BH-BH systems dominate event rates even in the
pessimistic ``High BH kick'' model (several events per year), wherein most of the systems
containing BHs are disrupted during the SN. Additionally, the BH-BH
mass distribution could have rich, observationally-accessible
structure (various lower limits and shapes) that encodes fine details
about stellar and binary evolution \citep[see,
  e.g.,][]{PSconstraints3-MassDistributionMethods-NearbyUniverse,massgap,2012ApJ...757...36K,chrisija}.
We note, however, that the crude binning in metallicity that we had to
undertake in order to limit computational costs may create artificial
sharp, narrow features in the mass distribution, which would merge
together into broader trends with a finer metallicity grid.

\cite{mennekens} point out that 
the detection rate of BH-BH systems may be reduced to zero due to the effects of 
intense stellar wind during the Red Supergiant and Luminous Blue Variable phases of BH progenitors. 
However, we have demonstrated that the \cite{mennekens} result is
a direct consequence of their assumption of no tidal interaction in close
binaries. If tides can efficiently transfer angular momentum from the orbit
into the companion spin, then it is expected that isolated binaries will
form close BH-BH systems. 

The criteria for the development of the CE phase may influence the merger and detection rates 
of all DCOs. \citep{woods} and \citep{ivanova2014} state that
the criterion for the stability of mass transfer sourced from the polytropic approximation
is much too strict. Therefore, the frequency of the CE may be overestimated. The CE
is a major mechanism for creating close binaries that coalesce within a Hubble time. 
The lack of CE events would, therefore, decrease the number of DCO mergers.
This would provide a reasonable pessimistic scenario for the lack of detections
of gravitational wave signals. A study of CE development criteria and its effect on the
formation of close BH-BH binaries is underway (Belczynski et al., in prep.).
However, an assumed rarity of CE systems would be difficult to reconcile
with observational evidence pointing to systems (for example V1309 Sco, V4332 Sgr, 
OGLE 2002-BLG-360 or CK Vul) which seem to have developed a CE (e.g., \cite{tylenda,martini99,
tylenda2013}). Additionally, massive X-ray binaries such as NGC300 X-1 or IC10 X-1 
are on close orbits with orbital periods $\sim 30$ hr, which have likely developed through a CE event.

Our study shows that detectable NS-NS systems are formed significantly
later in the history of the Universe than BH-BH and BH-NS systems. As
shown in Figs.~\ref{fig:FiducialResultDistributions:NSNS},
\ref{fig:FiducialResultDistributions:BHNS}, and
\ref{fig:FiducialResultDistributions:BHBH}, the birth times of NS-NS
systems cluster around $13$ Gyr after the Big Bang, while for the
other systems this is $1$ Gyr. This behavior might be
counter-intuitive, as the intrinsic distribution of time delays between
formation and merger for all types of DCOs falls off as $t_{\rm merger}^{-1}$,
barring exceptional circumstances \citep[e.g., near-solar metallicity
  BH-BH binaries,][]{dominik}.  Therefore, one might expect the
majority of detectable DCOs to be formed within the past $\sim$ Gyr as
is the case for NS-NS systems.  However, BH-BH systems are created
most efficiently in the lowest metallicity environments, and therefore
their formation rate is highest in the early Universe. The long
time-delay tail of these early systems dominates the subsequent
detection rate. The metallicity evolution is therefore a crucial
factor in predicting the detectable rate of DCOs.

We also find that including the merger and ringdown components of the
GW signal does not have a significant impact on the detection rates of
NS-NS systems. The full IMR calculations become important for higher
mass systems, and especially for BH-BH binaries.  The detection rates
for BH-BH systems increases by at least $20\%$, and typically by $\sim
50\%$, when using full IMR waveforms when compared to the PN inspiral
alone.

The detection rate of BH-BH systems is also sensitive to spin
effects. Extreme aligned spins increase the rates by a factor of $\sim
3$ when compared with the non-spinning case.

We used simplified criteria for detectability, considering an SNR
threshold of $8$ in a single detector as a proxy for the network
\citep[cf.~][]{2010CQGra..27q3001A}.  For reference, we also
considered a network SNR threshold of $10$, which is likely to be very 
optimistic, and $12$, which is more realistic \citep[cf.~][]{scenarios}, 
on a network of three detectors with aLIGO sensitivity.  The network SNR 
threshold of $12$ yields rates which are roughly comparable with rates 
computed using an SNR threshold of $8$ in a single aLIGO detector as proxy for the network.
The actual detection thresholds are a complicated function
of network configuration, the level and frequency of non-Gaussian,
non-stationary excursions in the noise, and search pipeline
sensitivity to different source types.  Therefore, our simple
thresholds are only meant to yield rough estimates of detection rates,
and the focus should be on relative rates for different source types
and model assumptions rather than absolute numbers.  Finally, we note
that the sensitivity of advanced detectors will gradually improve
during commissioning, and several years will pass before they reach
the sensitivity we have assumed above \citep[for an approximate time
  line, see][]{scenarios}.
  
The detection rates computed by assuming an SNR threshold of $8$ in a single aLIGO 
detector as proxy for the network allow for a direct comparison with the rate ranges 
compiled in \citep{2010CQGra..27q3001A}, which used the same detectability criterion.  
\citet{2010CQGra..27q3001A} incorporated a number of population synthesis studies and 
Galactic binary pulsar observations, but did not include some of the factors considered 
in the present study, such as cosmology and variations in metallicity distributions and star 
formation rates with redshift.  We find that our predicted detection rates for NS-NS and 
BH-BH binaries fall within the ranges given in \citep{2010CQGra..27q3001A} for all models 
and both metallicity distribution choices considered in the present work.  For BH-NS binaries, 
the same holds for all models and metallicity choices except for the high BH kick model, which 
yields BH-NS detection rates below the range quoted in \citep{2010CQGra..27q3001A}.

We note that uncertainties in waveform systematics and detection
criteria pale in comparison to uncertainties in stellar and binary
evolution. We consider the most important uncertainties to be the
progress and outcome of the CE phase, the SN explosion mechanism and
the magnitude of BH natal kicks. The four binary evolution models
discussed in this study explore these uncertainties, resulting in a
wide range of mass distributions and event rates.  Changing other
parameters such as the initial binary mass distribution or varying the
mass escaping the systems during mass transfer episodes would also
influence the resulting distributions and rates
\citep{2005ApJ...620..385O,2008ApJ...675..566O,roskb}.

The properties of the DCO populations produced in our various models
are sufficiently differentiated that it may be possible to constrain
or rule out some of the input physics based on observed populations.
For example, a lack of significant number of detections will disfavor the Optimistic CE model, in
which we allow for CE events with HG donors and thus find very high detection rates. 
This will indicate how (if at all) CE develops for HG stars.
If BH-BH systems are not detected far more
frequently than other DCO types, a likely explanation is that BHs
receive significant natal kicks disrupting their binaries. A detailed
comparison of detection rates with current LIGO upper limits can be
found in~\citet{comparison}.  As detections accumulate, a well measured
chirp mass distribution could allow us to distinguish between the
Rapid and Delayed SN engine models, which generate continuous and
gapped chirp mass distribution of DCOs, respectively. The number of
detections needed to distinguish between the Rapid and Delayed SN
engines will be discussed in future work (Dominik et al. 2014, in
preparation).

\acknowledgements

We thank a number of LIGO and Virgo collaboration colleagues,
particularly Thomas Dent, David Shoemaker, Stephen Fairhurst and Peter
Saulson, 
for advice on the manuscript.  We thank the N. Copernicus Astronomical
Centre in Warsaw, Poland, and the University of Texas at Brownsville,
for providing computational resources. The authors acknowledge the
Texas Advanced Computing Center (TACC) at The University of Texas at
Austin for providing computational resources. 
KB acknowledges support from a Polish Science Foundation “Master2013”
Subsidy, Polish NCN grant SONATA BIS 2, NASA Grant Number NNX09AV06A and NSF
Grant Number HRD 1242090 awarded to the Center for Gravitational Wave
Astronomy at U.T. Brownsville.
MD acknowledges support from the National Science Center grant
DEC-2011/01/N/ST9/00383.
EB acknowledges support
from National Science Foundation CAREER Grant PHY-1055103. 
ROS was supported by NSF award PHY-0970074 and the UWM Research Growth
Initiative.
DEH acknowledges support from National Science Foundation CAREER grant
PHY-1151836. He was also supported in part by the Kavli Institute for
Cosmological Physics at the University of Chicago through NSF grant
PHY-1125897 and an endowment from the Kavli Foundation and its founder
Fred Kavli.
TB was supported by the DPN/N176/VIRGO/2009 grant and the 
DEC-2013/01/ASPERA/ST9/00001 from the National Science Center, Poland.
FP was supported by STFC Grant No.~ST/L000342/1.
This work was supported in part by the National Science Foundation under 
Grant No. PHYS-1066293 and the hospitality of the Aspen Center for Physics (KB).
The study was also sponsored by the National Science Center grant Sonata Bis 2 
(DEC-2012/07/E/ST9/01360). 

\bibliographystyle{aa}
\bibliography{b1}

\begin{thebibliography}{107}
\expandafter\ifx\csname natexlab\endcsname\relax\def\natexlab#1{#1}\fi

\bibitem[{{Aasi} {et~al.}(2013{\natexlab{a}}){Aasi}, {Abadie}, {Abbott},
  {Abbott}, {Abbott}, {Abernathy}, {Accadia}, {Acernese}, {Adams}, {Adams}, \&
  et~al.}]{2013PhRvD..87b2002A}
{Aasi}, J., {Abadie}, J., {Abbott}, B.~P., {et~al.} 2013{\natexlab{a}}, \prd,
  87, 022002

\bibitem[{{Aasi} {et~al.}(2013{\natexlab{b}}){Aasi}, {Abadie}, {Abbott},
  {Abbott}, {Abbott}, {Abernathy}, {Accadia}, {Acernese}, \&
  et~al.}]{scenarios}
---. 2013{\natexlab{b}}, (arXiv:1304.0670)

\bibitem[{{Abadie} {et~al.}(2012){Abadie}, {Abbott}, {Abbott}, {Abbott},
  {Abernathy}, {Accadia}, {Acernese}, {Adams}, {Adhikari}, {Affeldt}, \&
  et~al.}]{2012PhRvD..85h2002A}
{Abadie}, J., {Abbott}, B.~P., {Abbott}, R., {et~al.} 2012, \prd, 85, 082002

\bibitem[{{Abadie} {et~al.}(2010){Abadie}, {Abbott}, {Abbott}, {Abernathy},
  {Accadia}, {Acernese}, {Adams}, {Adhikari}, {Ajith}, {Allen}, \&
  et~al.}]{2010CQGra..27q3001A}
---. 2010, Classical and Quantum Gravity, 27, 173001

\bibitem[{{Ajith}(2011)}]{ajithspin}
{Ajith}, P. 2011, \prd, 84, 084037

\bibitem[{{Ajith} \& {Bose}(2009)}]{ajithbose}
{Ajith}, P. \& {Bose}, S. 2009, \prd, 79, 084032

\bibitem[{{Ajith} {et~al.}(2011){Ajith}, {Hannam}, {Husa}, {Chen},
  {Br{\"u}gmann}, {Dorband}, {M{\"u}ller}, {Ohme}, {Pollney}, {Reisswig},
  {Santamar{\'{\i}}a}, \& {Seiler}}]{PhenomB}
{Ajith}, P., {Hannam}, M., {Husa}, S., {et~al.} 2011, Physical Review Letters,
  106, 241101

\bibitem[{{Amaro-Seoane} \& {Freitag}(2006)}]{Amaro:2006imbh}
{Amaro-Seoane}, P. \& {Freitag}, M. 2006, \apjl, 653, L53

\bibitem[{{Arroyo-Torres} {et~al.}(2013){Arroyo-Torres}, {Wittkowski},
  {Marcaide}, \& {Hauschildt}}]{yuscuti}
{Arroyo-Torres}, B., {Wittkowski}, M., {Marcaide}, J.~M., \& {Hauschildt},
  P.~H. 2013, \aap, 554, A76

\bibitem[{{Bailyn} {et~al.}(1998){Bailyn}, {Jain}, {Coppi}, \& {Orosz}}]{mg1}
{Bailyn}, C.~D., {Jain}, R.~K., {Coppi}, P., \& {Orosz}, J.~A. 1998, \apj, 499,
  367

\bibitem[{{Banerjee} {et~al.}(2010){Banerjee}, {Baumgardt}, \&
  {Kroupa}}]{banerjee}
{Banerjee}, S., {Baumgardt}, H., \& {Kroupa}, P. 2010, \mnras, 402, 371

\bibitem[{{Belczynski} {et~al.}(2014){Belczynski}, {Buonanno}, {Cantiello},
  {Fryer}, {Holz}, {Mandel}, {Miller}, \& {Walczak}}]{walczak}
{Belczynski}, K., {Buonanno}, A., {Cantiello}, M., {et~al.} 2014, \apj, 789,
  120

\bibitem[{{Belczynski} {et~al.}(2010){Belczynski}, {Dominik}, {Bulik},
  {O'Shaughnessy}, {Fryer}, \& {Holz}}]{nasza}
{Belczynski}, K., {Dominik}, M., {Bulik}, T., {et~al.} 2010, \apjl, 715, L138

\bibitem[{{Belczynski} {et~al.}(2012{\natexlab{a}}){Belczynski}, {Dominik},
  {Repetto}, {Holz}, \& {Fryer}}]{comparison}
{Belczynski}, K., {Dominik}, M., {Repetto}, S., {Holz}, D.~E., \& {Fryer},
  C.~L. 2012{\natexlab{a}}, ArXiv: 1208.0358

\bibitem[{{Belczynski} {et~al.}(2008{\natexlab{a}}){Belczynski}, {Kalogera},
  {Rasio}, {Taam}, {Zezas}, {Bulik}, {Maccarone}, \& {Ivanova}}]{startrack}
{Belczynski}, K., {Kalogera}, V., {Rasio}, F.~A., {et~al.} 2008{\natexlab{a}},
  \apjs, 174, 223

\bibitem[{{Belczynski} \& {Taam}(2008)}]{betaam2008}
{Belczynski}, K. \& {Taam}, R.~E. 2008, \apj, 685, 400

\bibitem[{{Belczynski} {et~al.}(2007){Belczynski}, {Taam}, {Kalogera}, {Rasio},
  \& {Bulik}}]{rarity}
{Belczynski}, K., {Taam}, R.~E., {Kalogera}, V., {Rasio}, F.~A., \& {Bulik}, T.
  2007, \apj, 662, 504

\bibitem[{{Belczynski} {et~al.}(2008{\natexlab{b}}){Belczynski}, {Taam},
  {Rantsiou}, \& {van der Sluys}}]{kabelspin}
{Belczynski}, K., {Taam}, R.~E., {Rantsiou}, E., \& {van der Sluys}, M.
  2008{\natexlab{b}}, \apj, 682, 474

\bibitem[{{Belczynski} {et~al.}(2012{\natexlab{b}}){Belczynski}, {Wiktorowicz},
  {Fryer}, {Holz}, \& {Kalogera}}]{massgap}
{Belczynski}, K., {Wiktorowicz}, G., {Fryer}, C.~L., {Holz}, D.~E., \&
  {Kalogera}, V. 2012{\natexlab{b}}, \apj, 757, 91

\bibitem[{{Bethe} \& {Brown}(1998)}]{bethe}
{Bethe}, H.~A. \& {Brown}, G.~E. 1998, \apj, 506, 780

\bibitem[{{BICEP2 Collaboration} {et~al.}(2014){BICEP2 Collaboration}, {Ade},
  {Aikin}, {Barkats}, {Benton}, {Bischoff}, {Bock}, {Brevik}, {Buder},
  {Bullock}, {Dowell}, {Duband}, {Filippini}, {Fliescher}, {Golwala},
  {Halpern}, {Hasselfield}, {Hildebrandt}, {Hilton}, {Hristov}, {Irwin},
  {Karkare}, {Kaufman}, {Keating}, {Kernasovskiy}, {Kovac}, {Kuo}, {Leitch},
  {Lueker}, {Mason}, {Netterfield}, {Nguyen}, {O'Brient}, {Ogburn}, {Orlando},
  {Pryke}, {Reintsema}, {Richter}, {Schwarz}, {Sheehy}, {Staniszewski},
  {Sudiwala}, {Teply}, {Tolan}, {Turner}, {Vieregg}, {Wong}, \&
  {Yoon}}]{2014arXiv1403.3985B}
{BICEP2 Collaboration}, {Ade}, P.~A.~R., {Aikin}, R.~W., {et~al.} 2014, arXiv:
  1403.3985

\bibitem[{{Bloom} {et~al.}(1999){Bloom}, {Sigurdsson}, \& {Pols}}]{bloom}
{Bloom}, J.~S., {Sigurdsson}, S., \& {Pols}, O.~R. 1999, \mnras, 305, 763

\bibitem[{{Capano} {et~al.}(2013){Capano}, {Pan}, \&
  {Buonanno}}]{Capano:2013raa}
{Capano}, C., {Pan}, Y., \& {Buonanno}, A. 2013, (arXiv:1311.1286)

\bibitem[{{Cutler} \& {Flanagan}(1994)}]{cutlerflanagan}
{Cutler}, C. \& {Flanagan}, {\'E}.~E. 1994, \prd, 49, 2658

\bibitem[{Damour {et~al.}(2011)Damour, Nagar, \& Trias}]{Damour:2010zb}
Damour, T., Nagar, A., \& Trias, M. 2011, Phys.Rev., D83, 024006

\bibitem[{{De Donder} \& {Vanbeveren}(1998)}]{dedonder}
{De Donder}, E. \& {Vanbeveren}, D. 1998, \aap, 333, 557

\bibitem[{{De Donder} \& {Vanbeveren}(2004{\natexlab{a}})}]{danny}
---. 2004{\natexlab{a}}, New Astronomy Review, 48, 861

\bibitem[{{De Donder} \& {Vanbeveren}(2004{\natexlab{b}})}]{ddv04}
---. 2004{\natexlab{b}}, \nar, 48, 861

\bibitem[{{Dewi} \& {Pols}(2003)}]{dewi}
{Dewi}, J.~D.~M. \& {Pols}, O.~R. 2003, \mnras, 344, 629

\bibitem[{{Dominik} {et~al.}(2012){Dominik}, {Belczynski}, {Fryer}, {Holz},
  {Berti}, {Bulik}, {Mandel}, \& {O'Shaughnessy}}]{dominik}
{Dominik}, M., {Belczynski}, K., {Fryer}, C., {et~al.} 2012, \apj, 759, 52

\bibitem[{{Dominik} {et~al.}(2013){Dominik}, {Belczynski}, {Fryer}, {Holz},
  {Berti}, {Bulik}, {Mandel}, \& {O'Shaughnessy}}]{dominik2}
---. 2013, \apj, 779, 72

\bibitem[{{Downing} {et~al.}(2010){Downing}, {Benacquista}, {Giersz}, \&
  {Spurzem}}]{downing}
{Downing}, J.~M.~B., {Benacquista}, M.~J., {Giersz}, M., \& {Spurzem}, R. 2010,
  \mnras, 407, 1946

\bibitem[{{Duch{\^e}ne} \& {Kraus}(2013)}]{duchene}
{Duch{\^e}ne}, G. \& {Kraus}, A. 2013, \araa, 51, 269

\bibitem[{{Finn}(1996)}]{finn96}
{Finn}, L.~S. 1996, \prd, 53, 2878

\bibitem[{{Finn} \& {Chernoff}(1993)}]{finnchernoff}
{Finn}, L.~S. \& {Chernoff}, D.~F. 1993, \prd, 47, 2198

\bibitem[{{Flanagan} \& {Hughes}(1998)}]{1998PhRvD..57.4535F}
{Flanagan}, {\'E}.~{\'E}. \& {Hughes}, S.~A. 1998, \prd, 57, 4535

\bibitem[{{Fontana} {et~al.}(2006){Fontana}, {Salimbeni}, {Grazian},
  {Giallongo}, {Pentericci}, {Nonino}, {Fontanot}, {Menci}, {Monaco},
  {Cristiani}, {Vanzella}, {de Santis}, \& {Gallozzi}}]{fontana}
{Fontana}, A., {Salimbeni}, S., {Grazian}, A., {et~al.} 2006, \aap, 459, 745

\bibitem[{{Fregeau} {et~al.}(2006){Fregeau}, {Larson}, {Miller},
  {O'Shaughnessy}, \& {Rasio}}]{Fregeau:2006}
{Fregeau}, J.~M., {Larson}, S.~L., {Miller}, M.~C., {O'Shaughnessy}, R., \&
  {Rasio}, F.~A. 2006, Astrophysical Journal Letters, 646, L135

\bibitem[{{Fryer} {et~al.}(2012){Fryer}, {Belczynski}, {Wiktorowicz},
  {Dominik}, {Kalogera}, \& {Holz}}]{chrisija}
{Fryer}, C.~L., {Belczynski}, K., {Wiktorowicz}, G., {et~al.} 2012, \apj, 749,
  91

\bibitem[{{Gerosa} {et~al.}(2013){Gerosa}, {Kesden}, {Berti}, {O'Shaughnessy},
  \& {Sperhake}}]{gerosa}
{Gerosa}, D., {Kesden}, M., {Berti}, E., {O'Shaughnessy}, R., \& {Sperhake}, U.
  2013, \prd, 87, 104028

\bibitem[{{Grindlay} {et~al.}(2006){Grindlay}, {Portegies Zwart}, \&
  {McMillan}}]{grin2006}
{Grindlay}, J., {Portegies Zwart}, S., \& {McMillan}, S. 2006, Nature Physics,
  2, 116

\bibitem[{{Grishchuk} {et~al.}(2001){Grishchuk}, {Lipunov}, {Postnov},
  {Prokhorov}, \& {Sathyaprakash}}]{grishchuk:2001}
{Grishchuk}, L.~P., {Lipunov}, V.~M., {Postnov}, K.~A., {Prokhorov}, M.~E., \&
  {Sathyaprakash}, B.~S. 2001, Physics Uspekhi, 44, 1

\bibitem[{{G{\"u}ltekin} {et~al.}(2004){G{\"u}ltekin}, {Miller}, \&
  {Hamilton}}]{gultekin}
{G{\"u}ltekin}, K., {Miller}, M.~C., \& {Hamilton}, D.~P. 2004, \apj, 616, 221

\bibitem[{{Harry, G.~M., for the LIGO Scientific
  Collaboration}(2010)}]{AdvLIGO}
{Harry, G.~M., for the LIGO Scientific Collaboration}. 2010, Classical and
  Quantum Gravity, 27, 084006

\bibitem[{{Heggie} \& {Giersz}(2014)}]{heggie2014}
{Heggie}, D.~C. \& {Giersz}, M. 2014, \mnras, 439, 2459

\bibitem[{{Hobbs} {et~al.}(2005){Hobbs}, {Lorimer}, {Lyne}, \&
  {Kramer}}]{hobbs}
{Hobbs}, G., {Lorimer}, D.~R., {Lyne}, A.~G., \& {Kramer}, M. 2005, \mnras,
  360, 974

\bibitem[{{Hogg}(1999)}]{hogg}
{Hogg}, D.~W. 1999, arXiv: astro-ph/9905116

\bibitem[{{Ivanova}(2015)}]{ivanova2014}
{Ivanova}, N. 2015, {Binary Evolution: Roche Lobe Overflow and Blue
  Stragglers}, ed. H.~M.~J. {Boffin}, G.~{Carraro}, \& G.~{Beccari}, 179

\bibitem[{{Ivanova} {et~al.}(2008){Ivanova}, {Heinke}, {Rasio}, {Belczynski},
  \& {Fregeau}}]{ivan}
{Ivanova}, N., {Heinke}, C.~O., {Rasio}, F.~A., {Belczynski}, K., \& {Fregeau},
  J.~M. 2008, \mnras, 386, 553

\bibitem[{{Ivanova} \& {Taam}(2004)}]{ivanovataam}
{Ivanova}, N. \& {Taam}, R.~E. 2004, \apj, 601, 1058

\bibitem[{{Kiminki} \& {Kobulnicky}(2012)}]{kiminki2}
{Kiminki}, D.~C. \& {Kobulnicky}, H.~A. 2012, \apj, 751, 4

\bibitem[{{Kiminki} {et~al.}(2012){Kiminki}, {Kobulnicky}, {Ewing}, {Bagley
  Kiminki}, {Lundquist}, {Alexander}, {Vargas-Alvarez}, {Choi}, \&
  {Henderson}}]{kiminki1}
{Kiminki}, D.~C., {Kobulnicky}, H.~A., {Ewing}, I., {et~al.} 2012, \apj, 747,
  41

\bibitem[{{Kreidberg} {et~al.}(2012){Kreidberg}, {Bailyn}, {Farr}, \&
  {Kalogera}}]{2012ApJ...757...36K}
{Kreidberg}, L., {Bailyn}, C.~D., {Farr}, W.~M., \& {Kalogera}, V. 2012, \apj,
  757, 36

\bibitem[{{Kroupa}(2014)}]{kroupa2014}
{Kroupa}, P. 2014, Astrophysics and Space Science Proceedings, 36, 335

\bibitem[{{Kulkarni} {et~al.}(1993){Kulkarni}, {Hut}, \& {McMillan}}]{kulkarni}
{Kulkarni}, S.~R., {Hut}, P., \& {McMillan}, S. 1993, \nat, 364, 421

\bibitem[{{Lipunov} {et~al.}(1997){Lipunov}, {Postnov}, \&
  {Prokhorov}}]{lipunov1997}
{Lipunov}, V.~M., {Postnov}, K.~A., \& {Prokhorov}, M.~E. 1997, \mnras, 288,
  245

\bibitem[{{Lyne} {et~al.}(2004){Lyne}, {Burgay}, {Kramer}, {Possenti},
  {Manchester}, {Camilo}, {McLaughlin}, {Lorimer}, {D'Amico}, {Joshi},
  {Reynolds}, \& {Freire}}]{Lyne:2004}
{Lyne}, A.~G., {Burgay}, M., {Kramer}, M., {et~al.} 2004, Science, 303, 1153

\bibitem[{{Mackey} {et~al.}(2008){Mackey}, {Wilkinson}, {Davies}, \&
  {Gilmore}}]{mackey}
{Mackey}, A.~D., {Wilkinson}, M.~I., {Davies}, M.~B., \& {Gilmore}, G.~F. 2008,
  \mnras, 386, 65

\bibitem[{{Mandel} \& {O'Shaughnessy}(2010)}]{MandelOShaughnessy:2010}
{Mandel}, I. \& {O'Shaughnessy}, R. 2010, Classical and Quantum Gravity, 27,
  114007

\bibitem[{{Marassi} {et~al.}(2011){Marassi}, {Schneider}, {Corvino}, {Ferrari},
  \& {Portegies Zwart}}]{seba}
{Marassi}, S., {Schneider}, R., {Corvino}, G., {Ferrari}, V., \& {Portegies
  Zwart}, S. 2011, \prd, 84, 124037

\bibitem[{{Martini} {et~al.}(1999){Martini}, {Wagner}, {Tomaney}, {Rich},
  {della Valle}, \& {Hauschildt}}]{martini99}
{Martini}, P., {Wagner}, R.~M., {Tomaney}, A., {et~al.} 1999, \aj, 118, 1034

\bibitem[{{Mennekens} \& {Vanbeveren}(2014)}]{mennekens}
{Mennekens}, N. \& {Vanbeveren}, D. 2014, \aap, 564, A134

\bibitem[{{Miller} \& {Lauburg}(2009)}]{MillerLauburg:2008}
{Miller}, M.~C. \& {Lauburg}, V.~M. 2009, \apj, 692, 917

\bibitem[{{Morscher} {et~al.}(2014){Morscher}, {Pattabiraman}, {Rodriguez},
  {Rasio}, \& {Umbreit}}]{morscher}
{Morscher}, M., {Pattabiraman}, B., {Rodriguez}, C., {Rasio}, F.~A., \&
  {Umbreit}, S. 2014, arXiv: 1409.0866

\bibitem[{{Nelemans} {et~al.}(2001){Nelemans}, {Yungelson}, \& {Portegies
  Zwart}}]{nele2001}
{Nelemans}, G., {Yungelson}, L.~R., \& {Portegies Zwart}, S.~F. 2001, \aap,
  375, 890

\bibitem[{{Nutzman} {et~al.}(2004){Nutzman}, {Kalogera}, {Finn}, {Hendrickson},
  \& {Belczynski}}]{nutzman}
{Nutzman}, P., {Kalogera}, V., {Finn}, L.~S., {Hendrickson}, C., \&
  {Belczynski}, K. 2004, \apj, 612, 364

\bibitem[{{O'Leary} {et~al.}(2009){O'Leary}, {Kocsis}, \& {Loeb}}]{OLeary:2008}
{O'Leary}, R.~M., {Kocsis}, B., \& {Loeb}, A. 2009, \mnras, 395, 2127

\bibitem[{{O'Leary} {et~al.}(2006){O'Leary}, {Rasio}, {Fregeau}, {Ivanova}, \&
  {O'Shaughnessy}}]{oleary}
{O'Leary}, R.~M., {Rasio}, F.~A., {Fregeau}, J.~M., {Ivanova}, N., \&
  {O'Shaughnessy}, R. 2006, \apj, 637, 937

\bibitem[{{O'Shaughnessy}(2013)}]{PSconstraints3-MassDistributionMethods-Nearb%
yUniverse}
{O'Shaughnessy}, R. 2013, PhRvD: 88, 084061

\bibitem[{{O'Shaughnessy} {et~al.}(2008){O'Shaughnessy}, {Belczynski}, \&
  {Kalogera}}]{2008ApJ...675..566O}
{O'Shaughnessy}, R., {Belczynski}, K., \& {Kalogera}, V. 2008, \apj, 675, 566

\bibitem[{{O'Shaughnessy} {et~al.}(2005){O'Shaughnessy}, {Kalogera}, \&
  {Belczynski}}]{2005ApJ...620..385O}
{O'Shaughnessy}, R., {Kalogera}, V., \& {Belczynski}, K. 2005, \apj, 620, 385

\bibitem[{{O'Shaughnessy} {et~al.}(2010{\natexlab{a}}){O'Shaughnessy},
  {Kalogera}, \& {Belczynski}}]{roskb}
---. 2010{\natexlab{a}}, \apj, 716, 615

\bibitem[{{O'Shaughnessy} {et~al.}(2010{\natexlab{b}}){O'Shaughnessy},
  {Vaishnav}, {Healy}, \& {Shoemaker}}]{2010PhRvD..82j4006O}
{O'Shaughnessy}, R., {Vaishnav}, B., {Healy}, J., \& {Shoemaker}, D.
  2010{\natexlab{b}}, \prd, 82, 104006

\bibitem[{{{\"O}zel} {et~al.}(2010){{\"O}zel}, {Psaltis}, {Narayan}, \&
  {McClintock}}]{mg2}
{{\"O}zel}, F., {Psaltis}, D., {Narayan}, R., \& {McClintock}, J.~E. 2010,
  \apj, 725, 1918

\bibitem[{{Pan} {et~al.}(2011){Pan}, {Buonanno}, {Boyle}, {Buchman}, {Kidder},
  {Pfeiffer}, \& {Scheel}}]{2011PhRvD..84l4052P}
{Pan}, Y., {Buonanno}, A., {Boyle}, M., {et~al.} 2011, \prd, 84, 124052

\bibitem[{{Pan} {et~al.}(2010){Pan}, {Buonanno}, {Buchman}, {Chu}, {Kidder},
  {Pfeiffer}, \& {Scheel}}]{eob}
{Pan}, Y., {Buonanno}, A., {Buchman}, L.~T., {et~al.} 2010, \prd, 81, 084041

\bibitem[{{Pannarale} {et~al.}(2013){Pannarale}, {Berti}, {Kyutoku}, \&
  {Shibata}}]{pannarale}
{Pannarale}, F., {Berti}, E., {Kyutoku}, K., \& {Shibata}, M. 2013, \prd, 88,
  084011

\bibitem[{{Panter} {et~al.}(2008){Panter}, {Jimenez}, {Heavens}, \&
  {Charlot}}]{panter}
{Panter}, B., {Jimenez}, R., {Heavens}, A.~F., \& {Charlot}, S. 2008, \mnras,
  391, 1117

\bibitem[{{Perets} \& {Kratter}(2012)}]{PeretsKratter:2012}
{Perets}, H.~B. \& {Kratter}, K.~M. 2012, \apj, 760, 99

\bibitem[{{Pfahl} {et~al.}(2005){Pfahl}, {Podsiadlowski}, \&
  {Rappaport}}]{pfahl}
{Pfahl}, E., {Podsiadlowski}, P., \& {Rappaport}, S. 2005, \apj, 628, 343

\bibitem[{{Poisson} \& {Will}(1995)}]{poissonwill}
{Poisson}, E. \& {Will}, C.~M. 1995, \prd, 52, 848

\bibitem[{{Portegies Zwart} \& {McMillan}(2000)}]{zwart2000}
{Portegies Zwart}, S.~F. \& {McMillan}, S.~L.~W. 2000, \apjl, 528, L17

\bibitem[{{Postnov} \& {Yungelson}(2006)}]{PostnovYungelson:2006}
{Postnov}, K.~A. \& {Yungelson}, L.~R. 2006, Living Reviews in Relativity, 9, 6

\bibitem[{{Repetto} \& {Nelemans}(2014)}]{serena}
{Repetto}, S. \& {Nelemans}, G. 2014, \mnras, 444, 542

\bibitem[{{Rodriguez} {et~al.}(2015){Rodriguez}, {Morscher}, {Pattabiraman},
  {Chatterjee}, {Haster}, \& {Rasio}}]{gcbhbh}
{Rodriguez}, C.~L., {Morscher}, M., {Pattabiraman}, B., {et~al.} 2015, arXiv:
  1505.00792

\bibitem[{{Sadowski} {et~al.}(2008){Sadowski}, {Belczynski}, {Bulik},
  {Ivanova}, {Rasio}, \& {O'Shaughnessy}}]{sadowski}
{Sadowski}, A., {Belczynski}, K., {Bulik}, T., {et~al.} 2008, \apj, 676, 1162

\bibitem[{{Santamar{\'{\i}}a} {et~al.}(2010){Santamar{\'{\i}}a}, {Ohme},
  {Ajith}, {Br{\"u}gmann}, {Dorband}, {Hannam}, {Husa}, {M{\"o}sta}, {Pollney},
  {Reisswig}, {Robinson}, {Seiler}, \& {Krishnan}}]{santamaria}
{Santamar{\'{\i}}a}, L., {Ohme}, F., {Ajith}, P., {et~al.} 2010, \prd, 82,
  064016

\bibitem[{{Schutz}(2011)}]{2011CQGra..28l5023S}
{Schutz}, B.~F. 2011, Classical and Quantum Gravity, 28, 125023

\bibitem[{{Shoemaker, D. for the LIGO Scientific Collaboration}(2010)}]{PSD:AL}
{Shoemaker, D. for the LIGO Scientific Collaboration}. 2010, {Advanced LIGO
  anticipated sensitivity curves}, https://dcc.ligo.org/LIGO-T0900288/public

\bibitem[{{Sigurdsson} \& {Hernquist}(1993)}]{sigurdsson}
{Sigurdsson}, S. \& {Hernquist}, L. 1993, \nat, 364, 423

\bibitem[{{Sippel} \& {Hurley}(2013)}]{sippel}
{Sippel}, A.~C. \& {Hurley}, J.~R. 2013, \mnras, 430, L30

\bibitem[{{Somiya}(2012)}]{KAGRA}
{Somiya}, K. 2012, Classical and Quantum Gravity, 29, 124007

\bibitem[{{Strolger} {et~al.}(2004){Strolger}, {Riess}, {Dahlen}, {Livio},
  {Panagia}, {Challis}, {Tonry}, {Filippenko}, {Chornock}, {Ferguson},
  {Koekemoer}, {Mobasher}, {Dickinson}, {Giavalisco}, {Casertano}, {Hook},
  {Blondin}, {Leibundgut}, {Nonino}, {Rosati}, {Spinrad}, {Steidel}, {Stern},
  {Garnavich}, {Matheson}, {Grogin}, {Hornschemeier}, {Kretchmer}, {Laidler},
  {Lee}, {Lucas}, {de Mello}, {Moustakas}, {Ravindranath}, {Richardson}, \&
  {Taylor}}]{strolger}
{Strolger}, L.-G., {Riess}, A.~G., {Dahlen}, T., {et~al.} 2004, \apj, 613, 200

\bibitem[{Taylor \& Weisberg(1989)}]{Taylor:1989}
Taylor, J.~H. \& Weisberg, J.~M. 1989, \apj, 345, 434

\bibitem[{Thorne(1974)}]{Thorne:1974ve}
Thorne, K.~S. 1974, Astrophys.J., 191, 507

\bibitem[{{Tsang}(2013)}]{Tsang:2013}
{Tsang}, D. 2013, \apj, 777, 103

\bibitem[{{Tylenda} {et~al.}(2011){Tylenda}, {Hajduk}, {Kami{\'n}ski},
  {Udalski}, {Soszy{\'n}ski}, {Szyma{\'n}ski}, {Kubiak}, {Pietrzy{\'n}ski},
  {Poleski}, {Wyrzykowski}, \& {Ulaczyk}}]{tylenda}
{Tylenda}, R., {Hajduk}, M., {Kami{\'n}ski}, T., {et~al.} 2011, \aap, 528, A114

\bibitem[{{Tylenda} {et~al.}(2013){Tylenda}, {Kami{\'n}ski}, {Udalski},
  {Soszy{\'n}ski}, {Poleski}, {Szyma{\'n}ski}, {Kubiak}, {Pietrzy{\'n}ski},
  {Koz{\l}owski}, {Pietrukowicz}, {Ulaczyk}, \& {Wyrzykowski}}]{tylenda2013}
{Tylenda}, R., {Kami{\'n}ski}, T., {Udalski}, A., {et~al.} 2013, \aap, 555, A16

\bibitem[{{Virgo Collaboration}(2009)}]{AdvVirgo}
{Virgo Collaboration}. 2009, Advanced Virgo Baseline Design, Virgo Technical
  Report VIR-0027A-09,
  https://tds.ego-gw.it/itf/tds/file.php?callFile=VIR-0027A-09.pdf

\bibitem[{{Voss} \& {Tauris}(2003{\natexlab{a}})}]{voss}
{Voss}, R. \& {Tauris}, T.~M. 2003{\natexlab{a}}, \mnras, 342, 1169

\bibitem[{{Voss} \& {Tauris}(2003{\natexlab{b}})}]{vosstauris}
---. 2003{\natexlab{b}}, \mnras, 342, 1169

\bibitem[{{Webbink}(1984)}]{webbink}
{Webbink}, R.~F. 1984, \apj, 277, 355

\bibitem[{{Wittkowski} {et~al.}(2012){Wittkowski}, {Hauschildt},
  {Arroyo-Torres}, \& {Marcaide}}]{wittkowski}
{Wittkowski}, M., {Hauschildt}, P.~H., {Arroyo-Torres}, B., \& {Marcaide},
  J.~M. 2012, \aap, 540, L12

\bibitem[{{Woods} \& {Ivanova}(2011)}]{woods}
{Woods}, T.~E. \& {Ivanova}, N. 2011, \apjl, 739, L48

\bibitem[{{Xu} \& {Li}(2010)}]{chlambda}
{Xu}, X.-J. \& {Li}, X.-D. 2010, \apj, 716, 114

\bibitem[{{Yuan} {et~al.}(2013){Yuan}, {Kewley}, \& {Richard}}]{yuan}
{Yuan}, T.-T., {Kewley}, L.~J., \& {Richard}, J. 2013, \apj, 763, 9

\bibitem[{{Yusof} {et~al.}(2013){Yusof}, {Hirschi}, {Meynet}, {Crowther},
  {Ekstr{\"o}m}, {Frischknecht}, {Georgy}, {Abu Kassim}, \&
  {Schnurr}}]{yusof2013}
{Yusof}, N., {Hirschi}, R., {Meynet}, G., {et~al.} 2013, \mnras, 433, 1114

\end{thebibliography}

\appendix

\section{Single and multidetector response}
The ``expected detection rate for GW detectors'' is a theorist's
idealization.  First and foremost, the event rate depends sensitively
on the (time-dependent) performance of instruments in development.
Furthermore, real GW searches employ complicated detection thresholds,
accounting for noise non-gaussianity and non-stationarity; for multiple instruments with
unequal power spectra; and for some search-dependent consistency
requirement across multiple detectors.
Rather than attempt realism, our idealizations provide a concrete,
reproducible filter to identify the number and (critically)
distribution of ``detectable'' binaries.

\subsection{Cumulative amplitude distribution for a single detector}

In a simple idealization, the detection threshold depends only on a
single detector's SNR.  Several authors have
characterized the response of a single GW detector to the angular
distribution of power for a GW source dominated by $(l,|m|)=(2,2)$
multipole radiation \citep{finnchernoff,finn96,roskb}.  This response depends on the 2-dimensional sky location $\Omega$, inclination $\iota$, and polarization $\psi$, and 
can be conveniently summarized by a projection parameter $w$ which is maximum ($w=1$) 
for a face-on, overhead source, and minimum ($w=0$) for sky locations and orientations where the detector has no response to the source.  The SNR, $\rho(\Omega,\psi,\iota)$, is equal to the maximum SNR of a face-on, overhead source at the same distance scaled by $w$, i.e., $\rho = w \rho_{\rm opt}$.
The cumulative distribution function for $w$ is $P(w)$:
\begin{eqnarray}
\label{eq:P}
P(w)&=& \int_{V}
          \frac{d\Omega}{4\pi}
          \frac{d\psi}{\pi}
          \frac{d\cos \iota }{2}         
\end{eqnarray}
where we integrate over the 4-dimensional angular integration volume, $V$, which is the set of all $\Omega,\iota,\psi$ such that the response exceeds $w$.   Our expression is identical to the cumulative distribution function $P(\Theta)$ defined by \cite{finnchernoff} and discussed also by
\cite{finn96}, but we use the variable $w=\Theta/4$ such that
$0<w<1$: \citep[see e.g.][]{roskb,walczak}. Note that $\langle w^2\rangle =
(2/5)^2$, therefore the optimal SNR at a given distance and the square root of
the angle-averaged signal power for a source at that distance ($\rho_{\rm ave}^2
\equiv \left<\rho^2\right>$) are related by $\rho_{\rm opt}=(5/2) \rho_{\rm ave}$.  Meanwhile, $\langle
w^3\rangle^{-1/3}\simeq 2.264$ is the factor commonly used to relate
volume-averaged distances to optimal detection distances, where $\left< w^3 \right>$ is the fraction of detectable sources within a sphere whose radius equals the at-threshold detection distance for an optimally located and oriented source; see, e.g., Eq.~(6) of \cite{roskb}.

Easily-interpolated tabulated results for $P(w)$ are available
online\footnote{Data files can be found online at the following URL:
  \url{http://www.phy.olemiss.edu/~berti/research.html}.}.  The
analytic approximation to this distribution function given by
\cite{finn96} is inadequate for our purposes; our tabulated results
follow from sampling the distribution numerically via a Monte Carlo
over $10^9$ binaries. We found that a good three-parameter fit to the
data is 
%
\be 
\label{eq:Pfit}
P(w)= a^{(n)}_{2} [(1-w/\alpha^{(n)})^{2}] + a^{(n)}_{4} [(1-w/\alpha^{(n)})^{4}] + a^{(n)}_{8} [(1-w/\alpha^{(n)})^{8}] + (1-a^{(n)}_{2}-a^{(n)}_{4}-a^{(n)}_{8})[(1-w/\alpha^{(n)})^{10}]\,,
\ee
where $(n)$ refers to the number of detectors in the network, $\alpha^{(n)}$ is the maximum value that $w$ can attain, so that $\alpha^{(1)}=1$ as $w$ is bounded between $0$ and $1$, and the coefficients are
$a^{(1)}_2 = 0.374222$, $a^{(1)}_4 = 2.04216$, and $a^{(1)}_8 =
-2.63948$. Notice that Eq.\,(\ref{eq:Pfit}) ensures that
$P(\alpha^{(1)})=0$ and $P(0)=1$. 

\subsection{Cumulative amplitude distribution for multiple detectors}
\label{ap:Details}
For a multidetector network $A$, a network SNR $\rho_A$
can always be defined.  Following an identical procedure as above, we
can define a cumulative distribution $P_A$ that generalizes
Eq.~(\ref{eq:P}).  
As before, $w=\rho/\rho_{\rm opt}$, but for multi-detector networks composed of
instruments with equal sensitivity, $\rho$ is the network SNR while $\rho_{\rm
  opt}$ is the single-detector SNR from an optimally-oriented binary directly
overhead that detector. 
For three identical instruments at the LIGO Hanford, Livingston, and Virgo sites, tabulated results for $P_A$ are available online at the URL listed
in the previous footnote; a good fit to the data has the form given in
Eq.~(\ref{eq:Pfit}), but now $0<w<1.4$, so
that $\alpha^{(3)}=1.4$. The coefficients we obtain are
%
$a^{(3)}_2 = 1.19549$, $a^{(3)}_4 = 1.61758$, and $a^{(3)}_8 =
-4.87024$.

\citet{2011CQGra..28l5023S} described a simple idealized model for the
sensitivity of multi-instrument networks.  This model is almost
equivalent to our own.  The two models differ in that
\citet{2011CQGra..28l5023S}, in his Eqs.(14)--(15), replaces $w^2$ by
an (unphysical) average of $w^2$ over polarization, then treats the
rms value of $w$ [i.e., $\left<w^2\right>^{1/2}$] as a substitute for
$w$ whenever $w$ appears.  Our results adopt no such simplifying
approximation.

\subsection{Higher harmonics}
Real GW sources produce multimodal radiation, with each mode providing
a distinct angular pattern.  For low-mass sources these higher
harmonics contribute little to the detector's response.  For high-mass
binaries with asymmetric mass ratios, higher harmonics can contribute
significantly to the observationally accessible signal \citep{Capano:2013raa}.
For nonspinning binaries of total mass $M<60 M_\odot$, and  with the smaller mass
 $>1.2M_\odot$, we
expect higher harmonics to increase the SNR $\rho$ by
less than a few percent, consistent with extrapolations derived using
PN waveforms. This expectation is supported by investigations carried out 
with a multimodal EOB IMR waveform
\citep{2011PhRvD..84l4052P}.
To a good approximation, the SNR $\rho$ and angular
distribution $P(w)$ can be approximated by the corresponding
expressions derived assuming purely quadrupolar, $(2,2)$-mode emission.

Higher harmonics can play a significant role if the mass distribution
extends to very high \emph{redshifted} mass.  At high mass, higher
harmonics contribute a greater fraction of the SNR, each
in a distinctive angular pattern; see \cite{2010PhRvD..82j4006O} for
illustrative results.
For aLIGO, systematic astrophysical uncertainties such as the BH spin
and mass have a significantly greater impact than the harmonic
content. These higher harmonics will be important for
third-generation interferometers, like the Einstein Telescope. This
will be investigated in future work.

\end{document}